\documentclass[a4paper,11pt]{article}
\pdfoutput=1 

\usepackage{jcappub} 

\usepackage[T1]{fontenc} 
\usepackage{fullpage,epsf}
\usepackage{amsmath}	
\usepackage{graphicx}	
\usepackage{amssymb}	
\usepackage{subcaption}

\title{\boldmath A Modal Approach to Constrain Inflation through Numerical Bispectra}

\usepackage{xcolor}

\author[a,1]{Bowei Zhang,\note{Corresponding author.}}
\author[a]{E.P.S Shellard,}
\author[a]{James R. Fergusson}

\affiliation[a]{Center for Theoretical Cosmology, DAMTP, University of Cambridge,\\Cambridge CB3 0WA, United Kingdom}

\emailAdd{bz287@cam.ac.uk}

\abstract{Constraining inflationary models with high precision bispectra across broad parameter ranges is a challenging task, requiring intensive computations at all stages, first, predicting the primordial inflation bispectrum from quantum field theory, secondly, projecting this forward with transfer functions to the late universe and, finally, comparing with the bispectrum extracted from the observational data and matching mock catalogues. Here, the longstanding separable \texttt{Modal} pipeline for constraining primordial bispectrum templates using WMAP and Planck CMB data \cite{Fergusson:2008ra,Fergusson:2009nv} has been supplemented by the more recently developed \texttt{Primodal} code \cite{Clarke:2020znk} to accurately calculate bispectra numerically from inflation models, showing great potential for enhanced computational efficiency;  \texttt{Primodal}  exploits the in-in separability of the tree-level in-in formalism, together with a separable mode-expansion technique to bypass the need for point-by-point bispectrum calculations.   Building upon this progress, we propose a bispectrum pipeline that systematically explores the parameter space of inflationary Lagrangians, numerically computing the tree-level bispectrum (and power spectrum) for each scenario  and comparing with the \texttt{Modal} bispectrum decompositions obtained from the Planck 2018 data. Our pipeline identifies and excludes disfavored scenarios through this analysis, providing direct constraints on the parameter space, the sound speed and other quantities from the surviving observationally viable scenarios.  This is preparatory work for a planned analysis using much higher-resolution CMB data from the Simons Observatory. To validate our pipeline, we perform a proof-of-concept analysis of the IR DBI inflation model, obtaining constraints of $c_s \geq 0.073$ for the sound speed and $\beta \leq 0.39$ for the parameter space, demonstrating the pipeline's accuracy and effectiveness.}

\begin{document}
\maketitle
\flushbottom

\section{Introduction}
Primordial non-Gaussianity (PNG) is a highly informative tool for exploring the early universe. Diverse inflation models predict PNG with varying amplitudes, shapes, and scale dependencies, offering means to differentiate among these models using observational data, a process initiated with WMAP CMB data \cite{WMAP:2003xez,WMAP:2008lyn,Senatore:2009gt} but which crossed a significant quantitative threshold with the Planck analysis \cite{Planck:2013wtn, Planck:2015zfm,Planck:2019kim} that was able to constrain a far greater range and variety of inflationary scenarios. The origin of PNG lies in the inherent non-linearity of inflationary processes that induce couplings between Fourier modes of the perturbations. This imprint typically manifests in higher-order correlators of observational data from the cosmic microwave background (CMB) and large-scale structure (LSS). In the context of the simplest inflation model -- featuring a single slow-rolling scalar field with a canonical kinetic term -- a calculable yet unobservable level of non-Gaussianity is predicted \cite{Maldacena:2002vr}. However, more complicated inflation models, encompassing self-interactions (represented as non-canonical kinetic terms) \cite{Alishahiha:2004eh, Arkani-Hamed:2003juy, Chen:2010xka, Burrage:2011hd}, interactions between multiple fields \cite{Byrnes:2010em}, as well as the presence of sharp or periodic features \cite{Chen:2006xjb,Adshead:2011jq,Flauger:2010ja}, can yield observable non-Gaussian correlations that were constrained by Planck \cite{Planck:2019kim}. To investigate the Non-Gaussian information of perturbation fields, we mostly rely on the three-point correlation function, whose Fourier counterpart is namely the bispectrum, as it represents the lowest-order statistic capable of distinguishing between Gaussian and non-Gaussian fields. While Gaussian-distributed perturbative fields are fully characterized by their two-point correlation function (or power spectrum), higher-order correlation functions carry the non-Gaussian information of primordial fluctuations. As such, the bispectrum stands as a pivotal observable in model-building, facilitating the exploration of the intriguing landscape of early universe cosmology.

Over the past two decades, considerable efforts have been devoted to classifying bispectra arising from inflationary models using standard templates \cite{Chen:2010xka}. For instance, single-field inflation with non-canonical kinetic terms can generate "equilateral" type of non-Gaussianity \cite{Alishahiha:2004eh,Arkani-Hamed:2003juy,Cheung:2007st} and "orthogonal" type of non-Gaussianity \cite{Senatore:2009gt}, while multi-field models can lead to "local" non-Gaussianity. When inflation models predict bispectra that can be well approximated by one standard template or a linear combination of such templates, constraints on these templates can be translated into constraints on model parameters. As a result, robust pipelines have been developed to constrain these standard templates using Planck data. The widely-used \textrm{KSW} estimator \cite{Komatsu:2003iq} accurately estimates the "nonlinearity parameter", $f_{NL}$ that quantifies the amplitude of non-Gaussianity, but it is limited to separable templates (separable templates refer to templates that can be factorized into separate functions of each momentum). These templates are, at best, approximate, including the
way they incorporate the spectral index. To broaden the range of models under investigation, the well-established \texttt{Modal} estimator \cite{Fergusson:2008ra,Fergusson:2009nv,Fergusson:2010dm,Fergusson:2014gea}, approximating templates as mode expansions over separable basis functions, allows for constraining non-separable templates (see also the related binned estimator \cite{Bucher:2009nm} and the modal KSW estimator ). However, the accuracy of the analysis may be compromised due to discrepancies between "real" bispectra and templates. Furthermore, with the high-resolution data from next-generation CMB experiments such as the Simons Observatory (SO) \cite{SimonsObservatory:2018koc} and CMB-S4 \cite{Abazajian:2019eic}, researchers naturally expect to explore inflation models directly by constraining precisely calculated bispectra beyond the realm of approximate standard templates. Therefore, a template-free pipeline is desirable, which utilizes the full bispectrum to directly confront inflation models with observational data.

The fundamental building block of the template-free analysis pipeline is an efficient method for the numerical computation of the primordial bispectrum, which is also a major challenge due to the computational complexity. Prior efforts in numerical calculations of inflationary bispectra, such as those in the \texttt{BINGO} code \cite{Hazra:2012yn}, Chen et al. \cite{Chen:2006xjb,Chen:2008wn}, the Transport Method \cite{Dias:2016rjq, Mulryne:2016mzv,Ronayne:2017qzn} and the Cosmological Flow Method \cite{Werth:2023pfl}, involve evaluating the bispectrum value configuration by configuration, resulting in a grid of points representing the primordial bispectrum. The first two works directly evaluate the in-in formalism \cite{Maldacena:2002vr} numerically, while the latter two utilise the Transport or the Cosmological Flow Method, solving a set of coupled differential equations in time to determine the correlator instead of doing the integral. There are two key issues that are common to these methods. Firstly, the configuration-by-configuration approach demands computationally expensive evaluations of the bispectrum values on a huge number of points to ensure the desired accuracy and resolution. Secondly, the output results cannot be directly compared with data using existing pipelines, such as the \textrm{KSW} estimator, for instance, because this requires a simple separable analytic expression for the shape. Even with the \texttt{Modal} estimator, evaluating the $f_{NL}$ of numerically computed bispectra across a broad parameter range necessitates decomposition with respect to many mode functions which entails considerable computational cost.     

Recently, a novel computational tool called \texttt{Primodal} was proposed and developed by Clarke et al. \cite{Clarke:2020znk} and \cite{Clarke:2022kvv}. This code utilizes the separability of the tree-level in-in formalism and incorporates the \texttt{Modal} philosophy to simplify the computation of primordial bispectra. Instead of evaluating the time-integral configuration-by-configuration (i.e.\ point-by-point in Fourier space), \texttt{Primodal} decomposes the curvature perturbation with respect to well-defined basis functions and solves much simpler time integrals of time-dependent coefficients mode-by-mode. The proof of concept for \texttt{Primodal} was presented in \cite{Funakoshi:2012ms} while a comprehensive development of the methodology, including the numerical setup and the definition of widely applicable basis functions, can be found in \cite{Clarke:2020znk}.  The \texttt{Primodal} pipeline demonstrates significantly improved computational efficiency by simplifying the integrands from the curvature perturbations to their mode decomposition, as well as by reducing the total number of integrals from that given by the number of configurations in $k$ to the number of basis functions. Notably, the i$\epsilon$ prescription is automatically taken into account in the \texttt{Primodal} approach because the highly oscillatory nature of mode functions at early times brings an additional suppressing factor to the integration \cite{Funakoshi:2012ms}, preventing the need for tricks to implement i$\epsilon$ numerically. Furthermore, the output bispectrum of \texttt{Primodal} adopts the form of a mode expansion based on a 3D separable basis. The format was designed for direct comparisons with observational data \cite{Clarke:2020znk} leveraging the CMB \texttt{Modal} pipeline or the related CMBEST estimator (a modal KSW approach introduced in \cite{Sohn:2023fte}).

In this work, we propose a conceptual pipeline that connects \texttt{Primodal} and \texttt{Modal} and facilitates a joint analysis of the power spectrum and the bispectrum to constrain inflation models from their Lagrangian directly. We employ the Infrared Dirac-Born-Infeld (IR DBI) inflation \cite{Alishahiha:2004eh,Chen:2005fe} as a test case, which is a well-known realisation of brane inflation and can generate large equilateral non-Gaussianities.  While we were completing this manuscript, we note that related work was placed on the arXiv in ref.~\cite{Philcox:2025bbo}, which exploits the same separable modal methodology to make tractable the confrontation between a general primordial bispectrum and observational data, specifically using a modal KSW estimator like CMBEST \cite{Sohn:2023fte}, though optimising the number of basis functions for particular bispectrum shapes. However, unlike our separable \texttt{Primodal} mode decomposition for evaluating the primordial bispectrum, this work uses the point-by-point CosmoFlow method \cite{Werth:2023pfl}. We also note advantages of the late-time Modal basis because it offers some degree of completeness and is more efficiently constructed from one-dimensional projections (i.e.\ not needing to project each 3D KSW mode): first, this allows us to reconstruct the full observed CMB bispectrum (TTT, TTE, TEE etc) at late times up to a given resolution \cite{Fergusson:2010dm, Planck:2019kim} thus combining primordial analysis with secondary effects and, secondly, we can explore large model parameter dependencies without recalculating our basis, e.g.\ as in the present numerical approach or  recent surveys of all the cosmological collider templates \cite{Suman:2025vuf} (see also \cite{Sohn:2024xzd}). 
 
The paper is organised as follows. In section \ref{section: Primodal}, we recap the in-in formalism for primordial non-Gaussianity and the underlying mathematical structure of the mode-decomposition method in inflation. In section \ref{section: model}, we present the model of the IR DBI inflation that we utilised to test our pipeline and we review the theoretical results of the PNG arising from this model. In section \ref{section: Modal}, we review the \texttt{Modal} philosophy and explain how to compare numerically predicted bispectra with CMB data with some conceptually simple basis conversions in \texttt{Modal} framework. In section \ref{sect:c_nl}, we define the consistency-level indicator $c_{NL}$ and introduce the concept of a template-free pipeline to constrain the inflationary parameter space. In section \ref{section:example}, we detail the methodology employed for validating our pipeline using the IR DBI model and present the numerical results of constraints on the parameter space and the sound speed of the inflation model. Finally, we discuss our current results by comparing them with the Planck 2018 constraints and summarise the outlook of future work in section \ref{section:conclusion}.

\label{sec:intro}

\section{Computing the In-In Correlator with the Modal-decomposition Approach}
\label{section: Primodal}
The separable \texttt{Modal} decomposition approach to calculating higher-order correlation functions entails exploiting separability to vastly improve efficiency and genericity, whether projecting forward primordial bispectra,  using transfer functions, or extracting observational bispectra from CMB or galaxy surveys \cite{Fergusson:2008ra,Fergusson:2009nv,Fergusson:2010ia}. A natural development was to apply this separable methodology to computing the primordial bispectrum during inflation, which was investigated and demonstrated for specific models in ref.~\cite{Funakoshi:2012ms}, and then a more broadly applicable implementation, together with a numerical code \texttt{Primodal}, was developed for general single-field models in ref.~\cite{Clarke:2020znk}. In this section, we briefly review the theoretical background of the inflationary power spectrum and bispectrum, then we demonstrate how the \texttt{Modal}-decomposition method can be applied to the numerical computation of primordial correlation functions.

\subsection{Curvature perturbation and The Primordial Power Spectrum}
In the uniform density gauge, where the inflaton field is unperturbed, the only perturbative quantity is the curvature perturbation $\zeta$ appearing in the metric. The second order action for the curvature perturbations is found \cite{Maldacena:2002vr} to be 
\begin{equation}
    S_2 = \int d^3xdt\frac{a \epsilon}{c_s^2}\left(a^2\dot{\zeta}^2-c_s^2(\partial \zeta)^2\right),
\label{2.1E1}
\end{equation}
where $a$ is the scaling factor, $\epsilon$ is the first order slow-roll parameter and $c_s$ is the sound speed.
By varying Eqn (\ref{2.1E1}), we could derive the equation of motion in the Fourier space
\begin{equation}
    \zeta_k''+(3-\epsilon+\eta-2\epsilon_s)\zeta_k'+\left(\frac{k c_s}{aH}\right)^2 \zeta_k = 0,
\label{2.1E2}
\end{equation}
where $\eta$ is the second order slow-roll parameter, $H$ is the Hubble rate and $\epsilon_s \equiv -\dot{c}_s/(Hc_s)$. Here, primes $'$ denote derivatives with respect to the number of e-folds.
Defining $v = -z\zeta$ where $z^2 = 2a^2\epsilon/c_s^2$, we obtain the familiar Mukhanov-Sasaki equation in conformal time,
\begin{equation}
    \frac{\partial^2v_k}{\partial \tau^2}+\left(c_s^2k^2-\frac{1}{z}\frac{d^2z}{d\tau^2}\right)v_k=0.
\label{2.1E3}
\end{equation}
In the most general cases, Eqn (\ref{2.1E2}) and Eqn (\ref{2.1E3}) need to be solved numerically. But we can take the slow varying parameters as constants to illustrate with a simple approximate solution in de-Sitter spacetime:
\begin{equation}
    \zeta_k(\tau) = \frac{H}{\sqrt{4\epsilon c_s k^3}}(1+i c_s k\tau)e^{-ic_s k\tau}\,,
\label{2.1E4}
\end{equation}
for which the tree-level primordial power spectrum at the end of inflation is found to be
\begin{equation}
    P_{\zeta}(k) = \left.\zeta_k\zeta_k^*\right|_{\tau = 0} \approx \frac{H^2}{4\epsilon c_s k^3}.
\label{2.1E5}
\end{equation}
Note that we shall actually use the numerically generated power spectrum with the appropriate spectral index (see Eqn.~(\ref{6E4})).

\subsection{The In-In Formalism and the Primordial Bispectrum}
If the field is non-Gaussian, it will be because higher-order correlators and loop corrections have arisen from the underlying interacting theory.  Generically, one-loop corrections of the power spectrum correspond to a quartic vertex or two cubic vertices, so they are typically weaker than the tree-level bispectrum which arises from a single cubic vertex.  For this reason, the bispectrum is well-motivated as the first observational target providing evidence for the non-Gaussianity of the Universe. 

To calculate the bispectrum, we need the interacting action of perturbations to at leasat the third order \cite{Maldacena:2002vr}. The standard approach to computing primordial correlation functions is the Schwinger--Keldysh formalism (also called the In-In formalism) as applied in ref.~\cite{Weinberg:2005vy}. The expectation value of the correlation function $O(t)$ in the interacting vacuum is given by
\begin{equation}
    \langle O(t) \rangle = \langle 0|\left[ \bar{\mathrm{T}} \mathrm{exp}\left(i \int_{-\infty(1+i\epsilon)}^{t}H_{I}(t')dt' \right)\right] O^{I}(t) \left[ \mathrm{T} \mathrm{exp}\left(-i \int_{-\infty(1-i\epsilon)}^{t}H_{I}(t'')dt'' \right)\right]|0\rangle,  
\label{2E1}
\end{equation}
where $|0\rangle$ denotes the vacuum of free theory and $\epsilon$ is a small positive real number which projects the integral onto the interacting vacuum. $H_{I}(t)$ denotes the interaction Hamiltonian, and $O^{I}$ is in the interaction picture. Using Eqn.~(\ref{2E1}), in Fourier space, the tree-level 3-point function of the primordial curvature perturbation field can be obtained by evaluating
\begin{equation}
    \langle \zeta_{k_1}(\tau)\zeta_{k_2}(\tau)\zeta_{k_3}(\tau) \rangle = -i\int_{-\infty(1-i\epsilon)}^{\tau}d\tau'a(\tau')\langle 0| \zeta_{k_1}(\tau)\zeta_{k_2}(\tau)\zeta_{k_3}(\tau)H_{(3)}(\tau')|0 \rangle + c.c.\,,
\label{2E2}
\end{equation}
where $\tau$ is the conformal time such that $d\tau = dt/a(t)$, and $H_{(3)}$ is the cubic interacting Hamiltonian. 

Following the convention used in the Planck CMB analysis \cite{Planck:2013wtn}, the primordial bispectrum $B(k_1, k_2, k_3)$ is defined by
\begin{equation}
    \lim_{\tau \rightarrow \infty}{\langle \zeta_{k_1}(\tau)\zeta_{k_2}(\tau)\zeta_{k_3}(\tau) \rangle} = (2\pi)^{3}\delta^{(3)}(\bold{k_1}+\bold{k_2}+\bold{k_3})\,B(k_1, k_2, k_3),
\label{2E3}
\end{equation}
and the dimensionless primordial shape function $S(k_1,k_2, k_3)$ is defined as 
\begin{equation}
   S(k_1,k_2, k_3) = (k_1k_2k_3)^2 B(k_1, k_2, k_3).
   \label{2E4}
\end{equation}

The primordial shape function is the quantity that we compute numerically and employ to constrain inflationary models. However, the term “shape function” emphasises only the geometric configuration of the bispectrum and can obscure the fact that the full amplitude of the numerically evaluated bispectrum also carries essential discriminatory power between different models and inflationary scenarios. To avoid this potential ambiguity, we use the term bispectrum throughout this paper. Whenever the term “bispectrum’’ appears in the following sections, it should be understood as referring to the bispectrum rescaled by the factor $(k_1k_2k_3)^2$.

\subsection{Analytical Templates and Template-based analysis}
Analytical computations of the in-in formalism have been performed for a significant number of inflation models, usually with suitable simplifying approximations. It is frequently noted that models inspired by different UV theories can generate bispectra with common shapes, which can be well-approximated by one of the equilateral, local and orthogonal shapes (or by a linear combination of them). These templates are explicitly given by \cite{Planck:2013wtn}:
\begin{align}
&S^{\mathrm{norm}}_{\mathrm{equil}} = \frac{(k_1k_2k_3)^2 (x_1+x_2-x_3)(x_2+x_3-x_1)(x_3+x_1-x_2)}{(x_1 x_2 x_3)^3},\label{2E9}
\\&S^{\mathrm{norm}}_{\mathrm{local}} = \frac{(k_1k_2k_3)^2}{3}\frac{(x_1^3+x_2^3+x_3^3)}{(x_1x_2x_3)^3},
\label{2E10}
\\&S^{\mathrm{norm}}_{\mathrm{ortho}} = \frac{(k_1k_2k_3)^2}{(x_1x_2x_3)^2} \left[3\left(\frac{x_1}{x_3}+\frac{x_2}{x_3}+\frac{x_1}{x_2}+\frac{x_3}{x_2}+\frac{x_2}{x_1}+\frac{x_3}{x_1}-\frac{x_1^2}{x_2x_3}-\frac{x_2^2}{x_1x_3}-\frac{x_3^2}{x_1x_2}\right)-8\right].
\label{2E11}
\end{align}
Here, in the expressions above, we use the sapproximation  
\begin{equation}
x_i\equiv k_*\left(\frac{k_i}{k_*}\right)^\frac{4-n_s}{3}\,,
\label{E50}
\end{equation}
where $k_* = 0.05\mathrm{Mpc}^{-1}$ is the conventional pivot scale for CMB analysis, and $n_s$ is the spectra index, which is introduced to capture the weak scale-dependence of non-Gaussianity from slow-roll dynamics \cite{Planck:2013wtn}. These templates are conventionally normalised by imposing $S^{\mathrm{norm}}(k_*,k_*,k_*) = 1$, although templates with strong scale dependence—such as those associated with feature or resonant models—require a modified normalisation scheme. We discuss alternative more robust definitions for normalisation later.  Improving observational constraints on these standard shapes (\ref{2E9}--\ref{2E11}) has been a longstanding objective in cosmology, as any robust detection would offer valuable insight into the underlying theory of inflation.

\subsection{In-In Separability and the {\large\texttt{Modal}} Decomposition Method}
By writing out the in--in formalism for the tree-level contact bispectrum explicitly, one finds that the primordial bispectra generated by any term in the cubic Hamiltonian can be expressed in a separable form~\cite{Funakoshi:2012ms}. More precisely, they can be written as a time integral of either a product, or a sum of products, of three functions depending separately on $k_1$, $k_2$, and $k_3$. As an illustrative example, the bispectrum arising from the operator $a(t)\omega(t)\dot{\zeta}(\partial\zeta)^2$ takes the form
\begin{align}
S_{\Dot{\zeta}(\partial\zeta)^2} & ~\propto~
i\bold{k_2}\cdot\bold{k_3} \int_{-\infty(1-i\epsilon)}^{\tau}d\tau'a(\tau')\,\omega(\tau')\{k_1^2\zeta_{k_1}(\tau)\,\zeta_{k_1}^{*'}(\tau')\}\{k_2^2\zeta_{k_2}(\tau)\,\zeta_{k_2}^{*}(\tau')\}\nonumber 
\\&~~~~~~~~~~~~~~~~~~~~~~~~~~~~~\times\{k_3^2\zeta_{k_3}(\tau)\,\zeta_{k_3}^{*}(\tau')\}+c.c+2\;\hbox{perms},
\label{2E5}
\end{align}
where $\bold{k_2}\cdot \bold{k_3} = (k_2^2+k_3^2-k_1^2)/2$.

Utilising this in-in separability, one could easily separate the time and wave-number dependence of the integrand by approximating functions like $k_i^2\zeta_{k_i}(\tau)\zeta_{k_i}^{*'}(\tau')$ and $k_i^2\zeta_{k_i}(\tau)\zeta_{k_i}^{*}(\tau')$ as mode expansions with respective to a one-dimensional orthonormal basis $q_{p}(k_i)$, in the form of 
\begin{equation}
   k_i^2\zeta_{k_i}(\tau)\zeta_{k_i}^{*'}(\tau') = \sum_p f_p(\tau')q_p(k_i), \qquad k_i^2\zeta_{k_i}(\tau)\zeta_{k_i}^{*}(\tau') = \sum_p g_p(\tau')q_p(k_i)  
   \label{2E6}
\end{equation}
One can move the time-independent mode functions outside the integral and evaluate the time integral of coefficients only. Afterwards, the full bispectrum can be reproduced as a three-dimensional mode expansion of the form:
\begin{equation}
   S(k_1,k_2,k_3) = \sum_{(i)}\sum_{pqr}\alpha_{pqr}^{(i)}q_p(k_1)q_q(k_2)q_r(k_3),
   \label{2E7}
\end{equation}
by adding the mode-functions back, where indices $(i)$ label different terms in the cubic Hamiltonian, and $\alpha$ are three-dimensional coefficients. For the $\Dot{\zeta}(\partial \zeta)^2$ term, $\alpha$ is of the form
\begin{equation}
   \alpha_{pqr}^{\Dot{\zeta}(\partial \zeta)^2} = \int d\tau'a(\tau')\omega(\tau')f_p(\tau')g_q(\tau')g_r(\tau')+2 \; \hbox{perms}.
   \label{2E8}
\end{equation}

Unlike traditional point-by-point codes, \texttt{Primodal} computes the primordial bispectrum mode by mode, evaluating a significantly smaller number of time integrals with simplified integrands \cite{Clarke:2020znk}. Moreover, the resulting bispectra are expressed directly as separable mode expansions, making them immediately suitable for comparison with data through the \texttt{Modal} CMB/LSS estimator.

\section{DBI Inflation}
\label{section: model}
In this work, we use the Infrared Dirac-Born-Infeld inflation model proposed in \cite{Chen:2004gc}\cite{Chen:2005ad}\cite{Bean:2007eh} to make a conceptual validation of the pipeline. The Lagrangian is given by
\begin{equation}
    S = \int d^4x\sqrt{-g}\left(-\frac{1}{f(\phi)}\left(\left(1+f(\phi)\partial_{\mu}\phi\partial^{\mu}\phi\right)^{{1}/{2}}-1\right)-V(\phi) \right)
    \label{4E1}
\end{equation}
where we take the canonical choices,
\begin{equation}
    f(\phi) = \frac{\lambda}{\phi^4},\quad V(\phi) = V_0-\frac{1}{2}m^2\phi^2,\quad m = \sqrt{\beta}H_i.
    \label{4E2}
\end{equation}
The constant $\lambda$ is proportional to the number of branes (inflatons) in the B-throat multiplied by the effective charge of the B-throat. Here, $V_0$ denotes the inflationary energy density and $\beta$ characterises the shape of the potential \cite{Bean:2007eh}. We will show that both the sound speed of inflation and the amplitude of the bispectrum depend primarily on the value of $\beta$. The Hubble parameter $H_i$ is evaluated at the initial time.

There are three principal motivations for selecting this model for our investigation. First, for suitable choices of parameters, it can yield a power spectrum compatible with the \textit{Planck} 2018 data. Second, the DBI model violates the assumptions underlying the single-field No-Go theorem and can therefore generate an observable level of primordial non-Gaussianity (PNG). Finally, as the DBI model is strongly motivated by string theory, adopting it as an example provides a clear illustration of how observational probes of the bispectrum can directly constrain the underlying parameters of fundamental theories.

\begin{figure}
    \centering
    \begin{subfigure}{0.85\textwidth}
        \centering
        \includegraphics[width=\columnwidth]{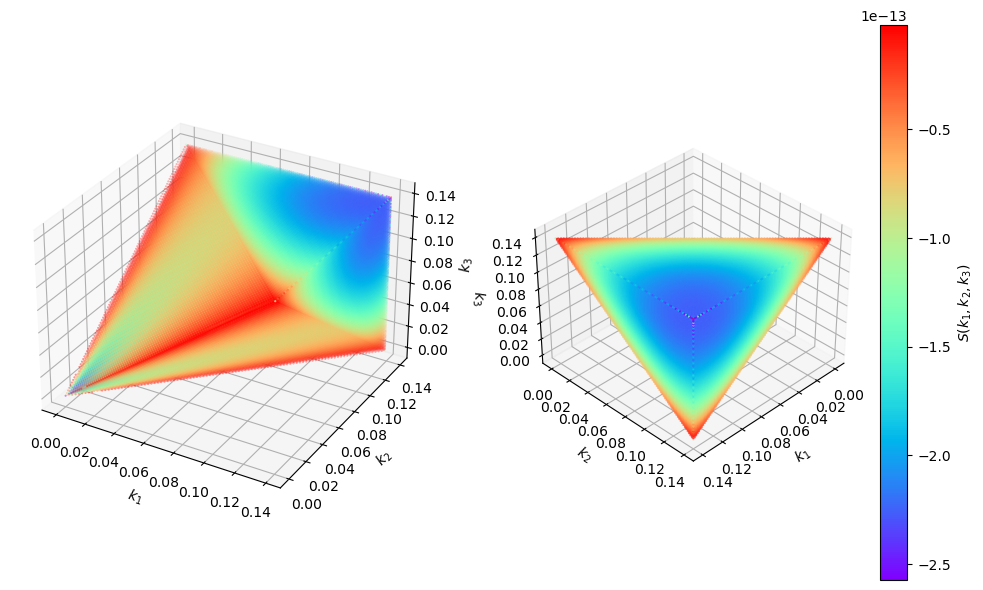}
        \caption{Numerical DBI shape}
        \label{3F1.1}
    \end{subfigure}
    \newline    \centering
    \newline    \centering
    \begin{subfigure}{0.75\textwidth}
        \centering
        \includegraphics[width=\columnwidth]{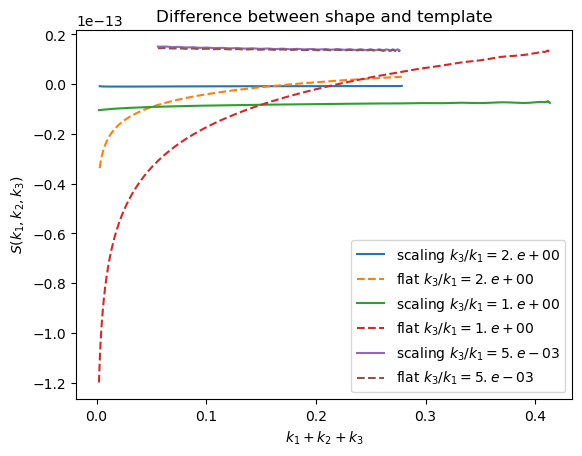}
        \caption{$S_{\mathrm{DBI}}^{\mathrm{shape}}-S_{\mathrm{DBI}}^{\mathrm{template}}$}
        \label{3F1.2}
    \end{subfigure}
    \newline
    \caption{(a) Three-dimensional plot of the numerical DBI bispectrum computed with \texttt{Primodal}; (b) The differences between the numerical DBI bispectrum and templates with scaling (solid lines) and without scaling (dashed line) at folded, equilateral and squeezed limits.}
    \label{3F1}
\end{figure}

There also exists a UV model of DBI inflation, originally proposed in ref.~\cite{Alishahiha:2004eh}. To leading order, the analytical calculation within perturbation theory yields the following template \cite{Alishahiha:2004eh}:
\begin{equation}
S_{\mathrm{DBI}}(k_1, k_2, k_3) = -\frac{35}{108}\left(\frac{1}{c_s^2}-1\right)6\left(\frac{H^2}{4\epsilon c_s}\right)^2\left(\frac{3}{5}\right)S_{\mathrm{DBI}}^{\mathrm{norm}}
\label{4E3}  
\end{equation}
where the sound speed $c_s$, the Hubble rate $H$ and the first-order slow-roll parameter $\epsilon$ are evaluated at horizon crossing.  Here, the normalised shape is
\begin{equation}
S_{\mathrm{DBI}}^{\mathrm{norm}} = \frac{1}{k_1k_2k_3}\frac{(-3/7)}{(k_1+k_2+k_3)^2}\left\{\sum_{i}k_i^5+\sum_{i\neq j}[2k_i^4k_j-3k_i^3k_j^2]+\sum_{i\neq j \neq l}[k_i^3k_jk_l-4k_i^2k_j^2k_l]  \right\},
\label{4E4}  
\end{equation}
which is strongly correlated with the standard equilateral template, given in eqn~(\ref{2E9}). This template can be modified to be more realistic by including the weak scale dependence encoded by the spectral index. The simplest way to achieve this is to multiply the template in eqn~(\ref{4E4}) with a prefactor 
\begin{equation}
\left(\frac{k_1 k_2 k_3}{k_*^3}\right)^{\frac{n_{\mathrm{NG}}}{3}}\qquad \hbox{or} \qquad \left(\frac{k_1+k_2+k_3}{3k_*}\right)^{n_{\mathrm{NG}}},
\label{4E6}  
\end{equation}
where $k_*$ is the pivot scale, and $n_{\mathrm{NG}} \equiv -2\Dot{c}_s/(H c_s)$, estimated at the horizon-crossing of pivot mode.  The accurate DBI bispectrum obtained numerically using \texttt{Primodal} is compared to the leading-order (scaled) template is illustrated in Figure~\ref{3F1}.
The theoretically predicted non-linearity parameter is
\begin{equation}
f_{\mathrm{NL}}^{\mathrm{DBI}} = -\frac{35}{108}\left(c_s^{-2} - 1\right),
\label{4E5}
\end{equation}
which offers a simplified target for the template-based analysis to directly link CMB constraints on $f_{\mathrm{NL}}$ to constraints on the sound speed $c_s$ of DBI inflation.

\section{Confronting Numerical Bispectra with CMB Data}
\label{section: Modal}

\begin{figure}
     \centering
     \includegraphics[width=\columnwidth]{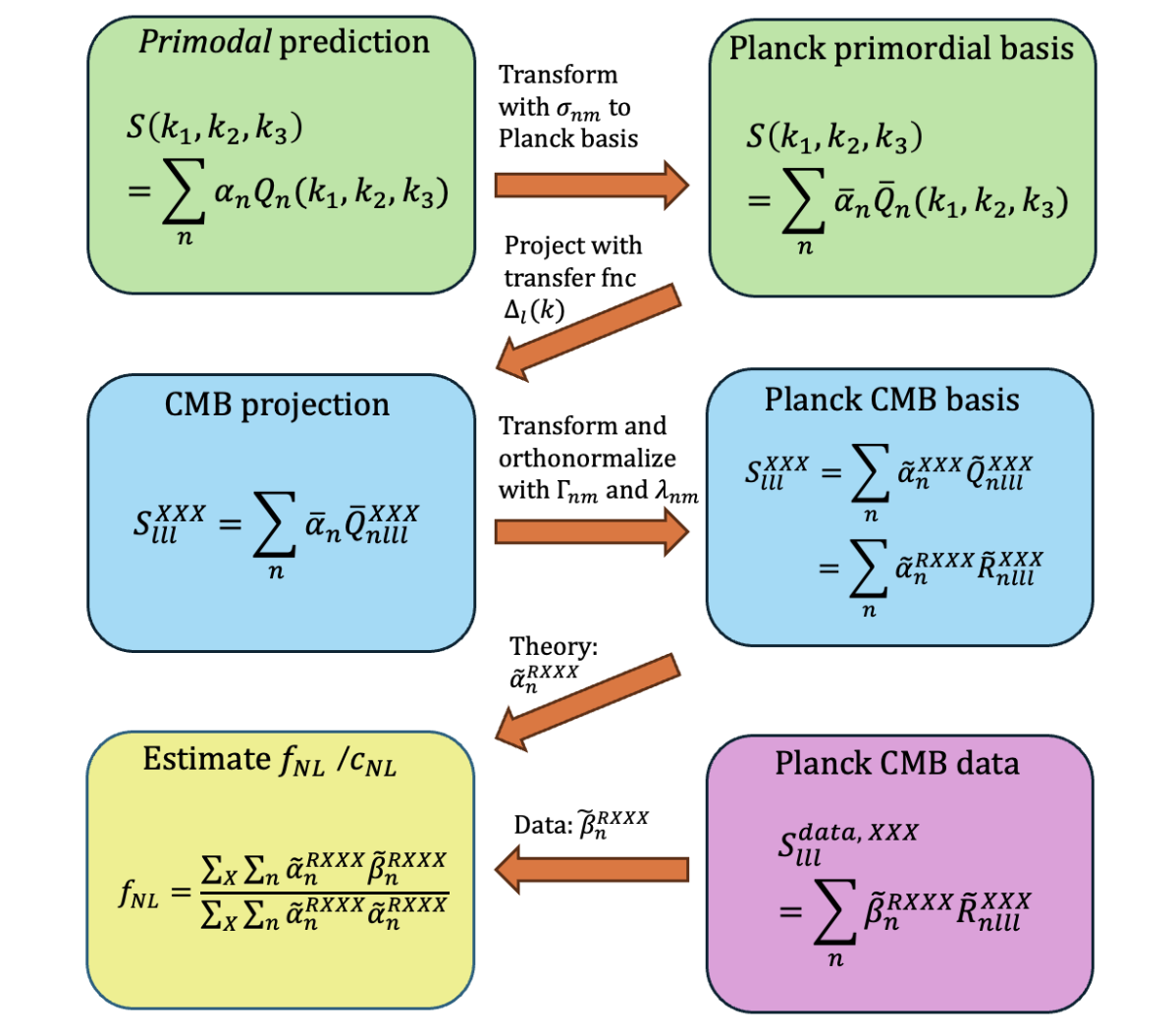}
     \caption{Graphic representation of the calculation and transformation steps involved in the \texttt{Modal} pipeline. The template-free analysis starts with the numerical computation of the In-In formalism with \texttt{Primodal} and ends with measurements of the consistency-level indicator $c_{\mathrm{NL}}$ defined in eqn~(\ref{E53}).  A simple template-based analysis starts with the mode-decomposition of analytical templates (the second block) and ends with the apparent "nonlinearity parameter" $f_{\mathrm{NL}}$.}
     \label{F4.1}
\end{figure}

This section outlines the procedures by which the numerically predicted bispectrum from \texttt{Primodal} can be directly compared with observational data from CMB experiments. In the \texttt{Modal} approach, bispectra are expressed as mode expansions with respect to different basis functions, with coefficients that are generally theory-dependent. The effectiveness of this methodology relies on all three-dimensional basis functions $Q_n(k_1, k_2, k_3)$ being expressible as separable products of one-dimensional basis functions. The sets of basis functions relevant to this work are listed below:

\begin{itemize}

\item \textbf{The {\large\texttt{Primodal}} basis} is denoted by $Q_{n}(k_1, k_2, k_3)$, which is used in the \texttt{Primodal} code for bispectra calculations over a separable cubic domain \cite{Clarke:2020znk}. 

\item \textbf{The Planck primordial basis} $\bar Q_{n}(k_1, k_2, k_3)$, which has been used to decompose the primordial bispectrum in the Planck analysis with the \texttt{Modal} pipeline, defined in wavenumber $k$-space over a tetrahedral-shaped domain, i.e.\ the {\it tetrapyd} \cite{Fergusson:2009nv}. 

\item \textbf{The projected Planck primordial basis} is denoted as $\bar Q_{n\;l_1 l_2 l_3}^{X_1 X_2 X_3}$, which is the direct projection of $\bar Q_{n}$ to $l$-space (angular multipole space) using linear transfer functions from the Einstein-Bolzmann equation, as in eqn~(\ref{E33}) \cite{Fergusson:2010dm}. Here, the upper indices $X$ can be one of $(T, E)$ denoting whether the multipoles are obtained for the temperature or E-mode polarisation CMB contributions \cite{Fergusson:2014gea}; the $X$'s are not tensor indices, so they do not contract. 

\item \textbf{The Planck CMB basis} $\widetilde Q_{n\;l_1 l_2 l_3}^{X_1X_2X_3}$ is the separable basis actually used to extract mode coefficients from the observational data and is defined on a multipole ($l$-space) tetrapyd, with its orthonormalisation  denoted by $\widetilde R_{n\;l_1 l_2 l_3}^{X_1X_2X_3}$\cite{Fergusson:2010dm,Fergusson:2014gea}.  It can be used to reconstruct the full CMB temperature and polarisation bispectrum \cite{Planck:2015zfm}. 
\end{itemize}
Comparing the numerical bispectrum from \texttt{Primodal} with the CMB data requires a series of basis transformations and projections, which is illustrated in Figure \ref{F4.1} and will be discussed in detail in this section.  The primary computational cost lies, first, in the initial CMB projection of the primordial separable basis, which allows prediction of the CMB bispectrum, and, secondly, in the separable extraction of the CMB bispectrum from the observed CMB map, together with large mock map catalogues that are generated using simulated CMB maps with the same masking, scanning and noise characteristics. Once the CMB modal projections and data extractions are completed, the primordial bispectrum is needed as input to search across model parameters and obtain constraints. 
\subsection{The {\large\texttt{Primodal}} Basis}
The fundamental purpose of \texttt{Primodal} in this pipeline is to compute general bispectra efficiently by using separable basis functions and express the outcomes as a mode decomposition.  The three-dimensional basis functions $Q_n$ are therefore constructed out of triplet products of normalised one-dimensional modes $q_p(k)$ as
\begin{equation}
Q_n(k_1,k_2,k_3)\equiv q_p(k_1)q_r(k_2)q_s(k_3),
\label{E27}
\end{equation}
where there is an appropriate mapping between integers $n$ and triplets $\{prs\}$, which are terminated at a given $p_{max}$. Once the basis is determined, it can be used to calculate the primordial bispectra of multiple classes of inflation models, and the bispectrum information of any specific scenario is encoded in the output expansion coefficients \cite{Clarke:2020znk}.

In the validation work discussed later, we use the so-called "Scaling Basis" to compute the bispectrum of the IR DBI inflation. The convergence of this basis has been studied for bispectra from general single-field inflation models with or without features in \cite{Clarke:2020znk}. The one-dimensional modes in the "scaling basis" with bespoke terms added to improve convergence are defined as 
\begin{equation}
 q_p(x)= 
\begin{cases}
    ln(k)/k,& \text{if } p = 0,\\
    1/k, & \text{if } p = 1,\\
    \mathcal{P}_{p-2}(\bar k), & \text{if } p > 1,
\end{cases}
\label{E5.28}
\end{equation}
where $\mathcal{P}$ are Legendre polynomials, and the first two mode functions are orthonormalised w.r.t the rest \cite{Clarke:2020znk}; these are especially relevant for local-type non-Gaussianity.   For $p > 2$, the argument of the mode functions is 
\begin{equation}
\bar k \equiv \frac{2k-(k_{\mathrm{max}}+k_{\mathrm{min}})}{k_{\mathrm{max}}-k_{\mathrm{min}}},
\label{E5.29}
\end{equation}
which lies in the range $[-1, 1]$, in agreement with the domain of Legendre polynomials.
\subsection{The Planck Primordial Basis}
In the Planck analysis, theoretical templates of the inflationary bispectra are decomposed over separable Planck Primordial Bases. In this report, we take the Fourier Planck primordial basis as an example, whose one-dimensional mode functions are defined as follows:
\begin{equation}
 q_i(x)= 
\begin{cases}
    \cos(i \pi x),& \text{if } i = \mathrm{even},\\
    \sin((i+1) \pi x),              & \text{if } i = \mathrm{old},
\end{cases}
\label{E28}
\end{equation}
where the argument $x$ is defined by $x = k/k_{max}$.

For a basis of order $p_{max}$, which includes one-dimensional basis functions with label $i$ ranging from $0$ to $p_{max}$, we substitute the 1-d mode functions with $i = p_{max}-1$ and $i = p_{max}$ by
\begin{align}
&\bar q_{p_{\mathrm{max}}-1}(x) = 1/x, \nonumber
\\&\bar q_{p_{\mathrm{max}}}(x) = x^2.
\label{E29}
\end{align}
These two augmented modes are especially effective at enabling the 3-d basis to capture any local behaviour in the primordial bispectra.
The 3-d mode functions are defined by taking the product of three 1-d mode functions and imposing all six symmetries by summing the relevant permutations together, which is given by \cite{Fergusson:2009nv}
\begin{align}
\bar Q_n(x,y,z) & \equiv 
\bar q_{\{ i}\bar q_j\bar q_{k\}} 
\label{E30}
\\&= \frac{1}{6}[\bar q_i(x)\bar q_j(y)\bar q_k(z)+\bar q_j(x)\bar q_k(y)\bar q_i(z)+\bar q_k(x)\bar q_i(y)\bar q_j(z)\nonumber \\&
\quad +\bar q_i(x)\bar q_k(y)\bar q_j(z)+\bar q_k(x)\bar q_j(y)\bar q_i(z)+\bar q_j(x)\bar q_i(y)\bar q_k(z)].
\label{E31}
\end{align}
We have specified a one-to-one mapping ordering the one-dimensional indices $ijk$ to a list labelled by $n$, which is similar to the definition of the \texttt{Primodal} three-dimensional basis functions. Given the symmetry over permutations, we also require $i\leq j\leq k$ in the $n \leftrightarrow ijk$ mapping.

In the validation work presented later, our one-dimensional basis is of order $p_{max} = 29$. The three-dimensional basis used for the \texttt{Modal} analysis \cite{Planck:2015zfm,Planck:2019kim} contains 2000 Fourier modes with $\sin$ and $\cos$ functions, and an augmented local mode: 
\begin{equation}
\bar Q_{1} = \frac{1}{3}\left(\frac{k_1^2}{k_2 k_3}+\frac{k_2^2}{k_3 k_1}+\frac{k_3^2}{k_1 k_2}\right). 
\label{E32}
\end{equation}
Some 3-d Fourier modes are deleted from the basis to remove degeneracy\cite{Planck:2015zfm}.

\subsection{From the {\large\texttt{Primodal}} basis to the Planck  primordial {\large\texttt{Modal}} basis}
In principle, any primordial bispectra $S$ can be written as mode expansions w.r.t.\ both the \texttt{Primodal} and Planck primordial basis:
\begin{equation}
S = \sum_{n} \alpha_{n} Q_{n} = \sum_{\bar n}\bar \alpha_{\bar n}\bar Q_{\bar n},
\label{E15}
\end{equation}
if a sufficiently large number of modes are included to ensure adequate convergence.
To compare the numerical predictions with the Planck data, our first step is transforming the \texttt{Primodal} coefficients $\alpha_{n}$ into the Planck primordial coefficients $\bar \alpha_{\bar n}$, where we note the complication that these basis functions are by construction separable, but not orthogonal. Nevertheless, this conversion between basis sets can be achieved using a linear transformation matrix $\sigma$ on the coefficients, 
\begin{equation}
 \bar \alpha_{\bar n} = \sum_{n} \sigma_{\bar n n}\alpha_{n}.
\label{E16}
\end{equation}
To evaluate the matrix $\sigma$, we firstly define the inner or dot product between $\bar Q_{\bar n}$ and $Q_{n}$ as
\begin{equation}
\omega_{\bar n m} = \langle\bar Q_{\bar n}, Q_{m}\rangle_k,
\qquad \hbox{or equivalently,} \qquad
\omega_{n \bar m}^{T} = \langle Q_{n}, \bar Q_{\bar m}\rangle_k,
\label{E18}
\end{equation}
where taking the inner product of two basis functions means integrating their product over the tetrapyd domain defined in eqn (\ref{E24}) \cite{Fergusson:2009nv}, as discussed below.  
We also define the inner product between mode functions in the Planck primordial basis as
\begin{equation}
\bar \gamma_{\bar n \bar m} = \langle\bar Q_{\bar n}, \bar Q_{\bar m}\rangle_k.
\label{E19}
\end{equation}
From eqn~(\ref{E15}) and eqn~(\ref{E16}), we obtain (summation assumed)
\begin{equation}
Q_{n} = \sigma^{T}_{n \bar m} \bar Q_{\bar m}\,.
\label{E20}
\end{equation}
Combining eqns~(\ref{E18}), (\ref{E19}) and (\ref{E20}), we find
\begin{equation}
\omega^{T}_{n \bar m} = \sigma^{T}_{n \bar l}\bar \gamma_{\bar l \bar m}\,.
\label{E21}
\end{equation}
Assuming a well-conditioned $\gamma_{\bar l \bar m}$, we multiply eqn~(\ref{E21}) on both sides by the inverse $\gamma^{-1}$ matrix to find
\begin{equation}
\omega^{T}_{n \bar m}\bar \gamma^{-1}_{\bar m \bar l} = \sigma^{T}_{n \bar l},
\label{E22}
\end{equation}
and the transformation matrix $\sigma$ is given by
\begin{equation}
\sigma_{\bar l n} = (\bar \gamma^{-1})^{T}_{\bar l \bar m}\omega_{\bar m n}\,.
\label{E23}
\end{equation}
Determining the $\sigma$ matrix requires the evaluation of the inner products between $\bar Q$ $\bar Q$, and $\bar Q$ $Q$ mode-by-mode. These computations can be done with high accuracy numerically or analytically and only need to be done once. With the $\sigma$ matrix, the basis transformation of the bispectrum coefficients becomes a simple linear matrix multiplication and can be done very efficiently for any bispectra decomposed in terms of the same basis.

Before ending this section, we provide some mathematical details about the integral required to take the inner product of two basis functions. As shown in \cite{Fergusson:2009nv}, the primordial bispectrum lives in the \textit{tetrapyd} domain, which is defined by 
\begin{equation}
k_1 \leq k_2+k_3,\quad k_2 \leq k_3+k_1,\quad k_3 \leq k_1+k_1\quad \textrm{and} \quad k_1,k_2,k_3\leq k_{\mathrm{max}}\,,
\label{E24}
\end{equation}
denoted as $\mathcal{V_{\mathcal{T}}}$. For an arbitrary function, $f(k_1,k_2,k_3)$, the integration over the tetrapyd is given explicitly by:
\begin{align}
\mathcal{T}[f] &= \int_{\mathcal{V_{\mathcal{T}}}}f(k_1,k_2,k_3)\;d\mathcal{V_{\mathcal{T}}}
\label{E25}
\\&= k_{max}^3\left\{\int_{0}^{\frac{1}{2}}\int_{y}^{1-y}\int_{x-y}^{x+y}F(x,y,z)dzdxdy+\int_{0}^{\frac{1}{2}}\int_{x}^{1-x}\int_{y-x}^{x+y}F(x,y,z)dzdydx\right. \nonumber 
 \\&\left.\quad+\int^{1}_{\frac{1}{2}}\int_{1-y}^{y}\int_{y-x}^{1}F(x,y,z)dzdxdy+\int^{1}_{\frac{1}{2}}\int_{1-x}^{x}\int_{x-y}^{1}F(x,y,z)dzdydx\right\},
\label{E26}
\end{align}
where we have made the transformation $x = k_1/k_{\mathrm{max}}$, $y = k_2/k_{\mathrm{max}}$, $z = k_3/k_{\mathrm{max}}$ with $F(x,y,z) = f(k_{\mathrm{max}}x,k_{\mathrm{max}}y,k_{\mathrm{max}}z)$. $x,y,z \in [0,1]$, define a normalised tetrapyd.  Therefore, the inner product between any two mode functions $\bar Q_{\bar n}$ and $Q_{m}$ from arbitrary basis sets can be evaluated from
\begin{equation}
\langle\bar Q_{\bar n}, Q_{m}\rangle_k = \mathcal{T}[ \bar Q_{\bar n}Q_{m}].
\label{E26.1}
\end{equation}

\subsection{From the Primordial Bispectrum to the CMB Bispectrum}
Now, we can express the numerically computed primordial bispectrum as a mode expansion w.r.t. the Planck primordial basis. In this subsection, we will discuss how the predicted primordial bispectrum is projected to late time "$l$-space", how it is converted to an expansion over a well-defined late-time basis and how the prediction is compared with observational data. The essential formulae align with those in \cite{Fergusson:2014gea} but are expressed differently in this overview.
\subsubsection{Projection to late time and conversion to the Planck CMB basis}
The Planck Primordial basis, $\bar Q_n(k_1, k_2, k_3)$, can be projected via the bispectrum transfer function $\Delta$ to the corresponding late-time CMB basis, $\bar Q_{n l_1 l_2 l_3}$. For temperature-only CMB maps, the projection is described by the solution of the linearised Einstein-Boltzmann equation and is given by
\begin{equation}
\bar Q_{n\;l_1 l_2 l_3}^{TTT} = \sqrt{\frac{v_{l_1}^2 v_{l_2}^2 v_{l_3}^2}{C_{l_1}^{TT}C_{l_2}^{TT} C_{l_3}^{TT}}} \int_{\mathcal{V_{\mathcal{T}}}} \bar Q_n(k_1, k_2, k_3) \Delta_{l_1, l_2, l_3}^{TTT}(k_1, k_2, k_3)d\mathcal{V_{\mathcal{T}}},
\label{E33}
\end{equation}
where the $\Delta_{l_1, l_2, l_3}^{TTT}(k_1, k_2, k_3)$ are CMB transfer functions with given cosmological parameters obtained from a suitable solver \cite{Lewis:1999bs,Diego_Blas_2011}.
For the primordial bispectrum described by a mode expansion over the Planck primordial basis $\bar Q_n(k_1, k_2, k_3)$, in the form of equation (\ref{E15}),
the projected CMB bispectrum then becomes
\begin{equation}
S_{l_1 l_2 l_3}^{X_1 X_2 X_3} = \sum_{n} \bar{\alpha}_{n} \bar{Q}_{n\;l_1 l_2 l_3}^{X_1 X_2 X_3}.
\label{E35}
\end{equation}
This projected early-time basis $\bar Q_{n l_1 l_2 l_3}$ can accurately represent the late-time primordial bispectrum $S_{l_1 l_2 l_3}^{X_1 X_2 X_3}$, but it is no longer a practical working basis for characterising the late-time CMB more generally. The projection involves integrations over many lines of sight combined with diffusive effects from the transfer functions, such as Silk damping, which introduces degeneracy between the smeared-out late-time modes.

For computational simplicity, a new late-time basis $\Tilde Q_{n\;l_1 l_2 l_3}^{X_1 X_2 X_3}$ is constructed directly in $l$-space in order to offer a more complete characterisation of the CMB bispectrum today, including the reconstruction of secondary bispectra like late-time ISW lensing. The new basis is described here as the Planck CMB basis   and is related to the projected early-time basis through transformation matrices $\Gamma_{np}^{X_1 X_2 X_3}$ defined as
\begin{equation}
\Gamma_{np}^{X_1 X_2 X_3} \equiv \sum_{r} (\Tilde{\gamma}^{X_1 X_2 X_3})_{nr}^{-1} \left\langle \Tilde{Q}_r^{X_1 X_2 X_3}, \quad \bar{Q}_p^{X_1 X_2 X_3}\right\rangle_l,
\label{E34}
\end{equation}
where $\Tilde{\gamma}^{X_1 X_2 X_3}_{nr}$ is defined in equation (\ref{E40}) below.
The expansion coefficients transform as
\begin{equation}
\Tilde \alpha_{n}^{X_1 X_2 X_3} = \sum_{p} \Gamma_{np}^{X_1 X_2 X_3} \bar \alpha_{p}\,,
\label{E36}
\end{equation}
and the theoretically predicted late-time bispectrum can, therefore, be represented by a mode expansion over the Planck CMB basis as follows:
\begin{equation}
S_{l_1 l_2 l_3}^{X_1 X_2 X_3, predict} = \sum_{n} \Tilde{\alpha}_{n}^{X_1 X_2 X_3} \Tilde{Q}_{n\;l_1 l_2 l_3}^{X_1 X_2 X_3}.
\label{E37}
\end{equation}
Now the late-time basis $\Tilde Q_{n\;l_1 l_2 l_3}^{X_1 X_2 X_3}$ is used to filter the CMB maps and extract the late-time CMB bispectrum from the Planck data for direct comparison with the projected bispectrum predictions \cite{Fergusson:2010dm,Planck:2013jfk, Planck:2019kim}.

\subsubsection{Orthonormalisation}
Mode functions of the Planck CMB basis $\Tilde{Q}_{n\;l_1 l_2 l_3}^{X_1 X_2 X_3}$ can be orthonormalised to form a new basis set $\Tilde{R}_{n\;l_1 l_2 l_3}^{X_1 X_2 X_3}$, which satisfies

\begin{equation}
\langle \Tilde{R}_{n\;l_1 l_2 l_3}^{X_1 X_2 X_3},\Tilde{R}_{p\;l_1 l_2 l_3}^{X_1 X_2 X_3} \rangle_{l} = \delta_{np}.
\label{E38}
\end{equation}
This orthonormalisation is done by taking the Cholesky decomposition of the associated $\gamma^{-1}$ defined as

\begin{equation}
(\Tilde \gamma^{X_1 X_2 X_2})_{np}^{-1} = \Tilde \lambda_{nr}^{X_1 X_2 X_2}(\Tilde \lambda_{rp}^{X_1 X_2 X_2})^T,
\label{E39}
\end{equation}
where 
\begin{equation}
\Tilde \gamma_{np}^{X_1 X_2 X_2} \equiv \langle \Tilde{Q}_{n\;l_1 l_2 l_3}^{X_1 X_2 X_3},\Tilde{Q}_{p\;l_1 l_2 l_3}^{X_1 X_2 X_3} \rangle_{l},
\label{E40}
\end{equation}
which is similar to equation (\ref{E19}) in $k$-space.

The relationship between the separable and orthonormal basis functions is 

\begin{equation}
\Tilde R_{n\;l_1 l_2 l_3}^{R X_1 X_2 X_3} = \Tilde \lambda_{np}^{X_1 X_2 X_2} \Tilde Q_{p\;l_1 l_2 l_3}^{X_1 X_2 X_3}.
\label{E41}
\end{equation}
Therefore, we can convert the original mode expansion of the predicted bispectrum in equation (\ref{E37}) to a new mode expansion in terms of the orthonormal basis,

\begin{equation}
S_{l_1 l_2 l_3}^{X_1 X_2 X_3, predict} = \sum_{n, X_1 X_2 X_3} \Tilde{\alpha}_{n}^{R X_1 X_2 X_3} \Tilde{R}_{n\;l_1 l_2 l_3}^{X_1 X_2 X_3},
\label{E42}
\end{equation}
and the coefficients w.r.t. orthonormal basis are related to those original by

\begin{equation}
\Tilde \alpha_{n}^{R X_1 X_2 X_3} = (\Tilde \lambda^{X_1 X_2 X_2})_{np}^{-1} \Tilde \alpha_{p}^{X_1 X_2 X_3}.
\label{E43}
\end{equation}
\subsection{Template-based $f_{\mathrm{NL}}$ Estimator}
The standard \texttt{Modal} Pipeline confronts normalised theoretical templates of the primordial bispectrum, defined by
\begin{equation}
    S(k_1, k_2, k_3) = f_{\mathrm{NL}}S^{\mathrm{norm}}(k_1,k_2,k_3),
\end{equation}
with the signal-to-noise CMB data in order to estimate $f_{\mathrm{NL}}$. 
Normalised templates are decomposed w.r.t the Planck Primordial basis by taking the inner products with mode functions \cite{Fergusson:2014gea}. The template is then projected to late-time $l$-space. After basis tranformation and orthonormalisation with equation (\ref{E36}) and equation (\ref{E43}), the theoretical templates are represented by coefficients $\Tilde{\alpha}_{n,\, \rm template}^{R X_1 X_2 X_3}$, in terms of orthonormal basis $\Tilde{R}_{n\; l_1 l_2 l_3}^{X_1X_2X_3}$.

Next, we return to the observational side. For simplicity writing down the temperature data only, under appropriate truncations, the $f_{\mathrm{NL}}$ estimator is given by
\begin{equation}
    f_{\mathrm{NL}} = \dfrac{1}{N} \sum_{l_jm_j} \dfrac{\mathcal{G}^{l_1l_2l_3}_{m_1m_2m_3} b_{l_1l_2l_3}}{C_{l_1}C_{l_2}C_{l_3}}  [a_{l_1m_1}a_{l_2m_2}a_{l_3m_3} - 3\langle a_{l_1m_1}a_{l_2m_2} \rangle^G a_{l_3m_3}].
    \label{4.4E1}
\end{equation}
where $a_{lm}$ are coefficients of the spherical harmonic decomposition of the CMB temperature anisotropies, $C_l$ is the power spectrum and $b_{l_1l_2l_3}$ is the reduced bispectrum from the normalised theoretical templates(See \cite{Fergusson:2009nv} for details).
By defining
\begin{equation}
    \bar{\Tilde{\beta}}_n =
    \int \mathrm{d}\Omega \Tilde{M}_{\{p}\Tilde{M}_{r}\Tilde{M}_{s\}}
    -
    \int \mathrm{d}\Omega 
    \left\langle\Tilde{M}^G_{\{p}\Tilde{M}^G_{r}
    \right\rangle
    \Tilde{M}_{s\}}
    \equiv
    \Tilde{\beta}_n^{\mathrm{cub}} - 3\Tilde{\beta}_n^{\mathrm{lin}},
    \label{4.4E2}
\end{equation}
where each $\tilde M_p$ is a realisation of the original CMB map filtered by the 1D basis function $q_p$,
\begin{equation}
    \Tilde{M}_p(\Omega) = \sum_{l,m} \Tilde{q}_p(l) 
    \dfrac{a_{lm}}{v_l\sqrt{C_l}}Y_{lm}(\Omega),
    \label{4.4E3}
\end{equation}
then the estimator reduces to
\begin{equation}
    f_{\mathrm{NL}} = \dfrac
    {\sum_n \tilde{\alpha}_n \tilde{\beta}_n}
    {\sum_{np} \tilde{\alpha}_n \tilde{\gamma}_{np}\tilde{\alpha}_p }
    = 
    \dfrac
    {\sum_n \tilde{\alpha}_n^{\mathcal{R}} \tilde{\beta}_n^{\mathcal{R}}}
    {\sum_{n} \tilde{\alpha}^{\mathcal{R}}_n \tilde{\alpha}^{\mathcal{R}}_n}.
    \label{4.4E4}
\end{equation}
Here, $\beta$ can be interpreted as the coefficients of the data bispectrum. Including polarisation and after orthonormalisation, we can reconstruct each component of the CMB bispectrum ($XXX = TTT, TTE, TEE$ etc) as the following
\begin{equation}
S_{l_1 l_2 l_3}^{X_1 X_2 X_3, \mathrm{data}} = \sum_{n, X_1 X_2 X_3} \Tilde{\beta}_{n}^{R X_1 X_2 X_3} \Tilde{R}_{n\;l_1 l_2 l_3}^{X_1 X_2 X_3}.
\label{E44.2}
\end{equation}
The ability to reconstruct late-time bispectrum components is an invaluable capability of the complete \texttt{Modal} approach because it allows for the simultaneous analysis of both primordial and late-time secondary bispectrum contributions (with the latter, like ISW-lensing, typically contaminating inflationary signals).  In the orthogonal basis, the estimator then finally becomes a simple dot product between coefficients for the observed and predicted bispectra:\cite{Fergusson:2014gea}
\begin{align}
f_{\mathrm{NL}} & = \frac{\Sigma_{X_i}\Sigma_{n}\Tilde{\alpha}_{n,\,\rm template}^{R X_1 X_2 X_3}\,\tilde{\beta}_n^{R X_1 X_2 X_3}}{\Sigma_{X_i}\Sigma_{n}(\tilde{\alpha}_{n,\, \rm template}^{R X_1 X_2 X_3})^2}
\label{E45}
\\&= \frac{\Sigma_{n}\alpha_n^{RTTT}\beta_n^{RTTT}+3\Sigma_{n}\alpha_n^{RTTE}\beta_n^{RTTE}+3\Sigma_{n}\alpha_n^{RTEE}\beta_n^{RTEE}+\Sigma_{n}\alpha_n^{REEE}\beta_n^{REEE}}{\Sigma_{n}(\alpha_n^{RTTT})^2+3\Sigma_{n}(\alpha_n^{RTTE})^2+3\Sigma_{n}(\alpha_n^{RTEE})^2+\Sigma_{n}(\alpha_n^{REEE})^2},
\label{E46}
\end{align}
where $\tilde{\alpha}_n$ are the late-time mode expansion coefficients for the predicted primordial bispectra (in the second equality, we drop the $\sim$ on $\alpha$ and $\beta$ for simplicity).

In the Planck convention, standard templates are of the form:
\begin{equation}
S_{\Phi}(k_1,k_2,k_3) = 6A_{\Phi}^2 f_{\mathrm{NL}}S^{\mathrm{norm}}(k_1,k_2,k_3).
\label{E46.2}
\end{equation}
The factor of $6A_{\Phi}^2$, for which $A_{\Phi}$ is the amplitude of the power spectrum of the Bardeen potential ($\Phi$ = 3/5$\zeta$ at super-horizon), has already been extracted from the data when people reconstruct the signal-to-noise CMB bispectrum.

Normalised templates can be phenomenological templates for equilateral, local, and orthogonal shapes in Eqn (\ref{2E9}) (\ref{2E10}) and (\ref{2E11}), and can also be model-specific templates, such as Eqn (\ref{4E4}) for the DBI inflation.

The errors of $f_{\mathrm{NL}}$ are estimated from the 160 FFP10 Planck CMB simulation maps \cite{Planck:2019kim}. Once the central value and error of $f_{\mathrm{NL}}$ are obtained for a specific template from an inflation model, constraints on this model can be derived from theoretical relationships between $f_{\mathrm{NL}}$ and model parameters. For example, for IR DBI inflation, which is introduced in section \ref{section: model}, people find $c_s \approx (\beta N/3)^{-1}$ \cite{Chen:2005fe}. Further, the theoretical expression of amplitude, equation (\ref{4E5}), becomes 
\begin{equation}
f_{\mathrm{NL}}^{\mathrm{DBI}} =  -\frac{35}{108}\left[\left( \frac{\beta^2 N^2}{9}-1 \right) \right].
\label{E51}
\end{equation}
With this relationship, the Planck 2013 collaboration \cite{Planck:2013wtn} found $\beta \leq 0.7$ at $95\%$ confidence level from $f_{\mathrm{NL}}^{\mathrm{DBI}} = 11\pm 69$, with CMB temperature data. 

\section{Template-free Pipeline and $c_{\mathrm{NL}}$}
\label{sect:c_nl}
As discussed in the last section, current data analysis for the PNG from CMB data relies on approximate analytic templates, as well as the analytic relationships between physical quantities and $f_{\mathrm{NL}}$. Therefore, assumptions made in such an analytical study may increase the inaccuracy of the data analysis, especially for more complicated models with scale-dependent physics. In our \texttt{Primodal}+\texttt{Modal} pipeline, instead of confronting normalised theoretical templates to the data bispectrum, we directly compare numerically calculated bispectra from \texttt{Primodal} with the data bispectrum. By doing so, we can investigate not only the shape but the amplitude of the bispectrum from any scenarios of one specific inflation model, and the consistency level between prediction and data directly tells us whether current observational results favour this scenario/model or not. To achieve this purpose, we need to define an indicator to determine how well the predicted bispectrum fits the data. We directly borrow the idea of the $f_{\mathrm{NL}}$ estimator in equation (\ref{E46}) to define the consistency-level indicator, $c_{\mathrm{NL}}$, the estimator of which is given by
\begin{align}
c_{NL} & = \frac{\Sigma_{X_i}\Sigma_{n}\alpha_n^{R X_1 X_2 X_3}\beta_n^{R X_1 X_2 X_3}}{\Sigma_{X_i}\Sigma_{n}(\alpha_n^{R X_1 X_2 X_3})^2}
\label{E52}
\\&= \frac{\Sigma_{n}\alpha_n^{RTTT}\beta_n^{RTTT}+3\Sigma_{n}\alpha_n^{RTTE}\beta_n^{RTTE}+3\Sigma_{n}\alpha_n^{RTEE}\beta_n^{RTEE}+\Sigma_{n}\alpha_n^{REEE}\beta_n^{REEE}}{\Sigma_{n}(\alpha_n^{RTTT})^2+3\Sigma_{n}(\alpha_n^{RTTE})^2+3\Sigma_{n}(\alpha_n^{RTEE})^2+\Sigma_{n}(\alpha_n^{REEE})^2}.
\label{E53}
\end{align}
for the T+E data, and
\begin{equation}
c_{\mathrm{NL}} = \frac{\Sigma_n \alpha_n^{RTTT}\beta_n^{RTTT}}{\Sigma_{n}(\alpha_n^{RTTT})^2}
\label{E3.1}
\end{equation}
for temperature data only.
This time, $\alpha_n^{RXXX}$s are coefficients of the numerical bispectrum calculated by \texttt{Primodal} (projected to $l$-space and transformed to orthonormal basis), which has not been scaled to unit like the templates, therefore both its shape and its amplitude are physically meaningful. We divide these $\alpha_n^{RXXX}$ coefficients by $6\left(\frac{3}{5}\right)A_{\zeta}^2$ to remove the amplitude of the power spectrum (the additonal factor of $3/5$ corresponds to the difference between the curvature perturbation and the Bardeen potential). Intuitively, if the predicted bispectrum is consistent with the data, we expect to have $\alpha_n \approx \beta_n$ mode-by-mode, therefore $c_{\mathrm{NL}} \approx 1$. 

\subsection{The Constraining Criteria}
For any specific scenario arising from a given inflationary model, one can estimate its value of $c_{\mathrm{NL}}$ from the numerically computed bispectrum and CMB data using equation~(\ref{E52}), and subsequently compare the measurement with the theoretical benchmark value $c_{\mathrm{NL}} = 1$. A scenario is deemed disfavoured by the data if $c_{\mathrm{NL}} = 1$ lies outside the $2\sigma$ confidence interval of the estimated value; more concretely, if  
\[
c_{\mathrm{NL}} + 2\sigma < 1 \quad \text{or} \quad c_{\mathrm{NL}} - 2\sigma > 1,
\]  
where $\sigma$ is the standard deviation of $c_{\mathrm{NL}}$ obtained from 160 Gaussian simulations, evaluated using the same methodology as for $f_{\mathrm{NL}}$. In summary, we rule out scenarios or models by excluding the condition $c_{\mathrm{NL}} = 1$. We express this bispectrum-based exclusion criterion in a compact form as follows:

\begin{equation}
|c_{\mathrm{NL}}-1| < 2 \sigma.
\label{E3.2}
\end{equation}

By introducing the $c_{\mathrm{NL}}$ parameter, constraining the inflationary parameter space using the bispectrum becomes straightforward and conceptually simple: one may scan across the parameter space to identify all scenarios that are not ruled out by equation~(\ref{E3.2}). The boundary of the remaining viable region can then be naturally interpreted as the resulting constraint.

\subsection{The Discovery Criteria}
However, detecting $c_{\mathrm{NL}} \approx 1$ is insufficient to declare the discovery of a good model/scenario; people also need to check the significance: $|\frac{c_{\mathrm{NL}}}{\sigma}|$ to distinguish non-Gaussian signature from Gaussian noise. In statistics, a result is typically regarded as significant only if it lies outside at least the $3\sigma$ region of the Gaussian probability distribution associated with the null hypothesis. In our convention, this corresponds to  
\begin{equation}
\left|\frac{c_{\mathrm{NL}}}{\sigma}\right| > 3.
\label{E3.3}
\end{equation}
and we adopt this as the discovery criterion for the primary search for primordial non-Gaussianity.

For models exhibiting strong scale dependence, such as those with primordial features or resonant non-Gaussianity---different parameter choices (e.g.\ the position of a sharp feature or the frequency of a resonant feature) can generate nearly linearly independent bispectra. In such cases, a significance exceeding $3\sigma$ may be achieved purely due to statistical fluctuations. When this occurs, one may either impose a more stringent discovery threshold, for example $\left|\frac{c_{\mathrm{NL}}}{\sigma}\right| > 5$, following the convention in particle physics, or explicitly account for the look-elsewhere effect and report an adjusted signal-to-noise ratio (SNR). Notable examples include \cite{Sohn:2024xzd} and \cite{Suman:2025vuf}, both of which employ the \texttt{Modal} approach in the context of cosmological collider models.

\subsection{Comparing $f_{\mathrm{NL}}$ and $c_{\mathrm{NL}}$}
In this subsection, we discuss the relationship and difference between the $f_{\mathrm{NL}}$ and $c_{\mathrm{NL}}$, and then highlight how template-free $c_{\mathrm{NL}}$ analysis incorporates the amplitudes of non-Gaussianities.

Suppose that we have the theoretically predicted bispectrum $S_{\mathrm{th}}^{XXX} = \sum_n \Tilde{\alpha}_n^{XXX}\tilde{R}_{n\;lll}^{XXX}$ and the CMB data bispectrum $S_{\mathrm{data}}^{XXX} = \sum_n \Tilde{\beta}_n^{XXX}\tilde{R}_{n\;lll}^{XXX}$, where $\tilde{\alpha}_n^{XXX}$ and $\tilde{\beta}_n^{XXX}$ are coefficients w.r.t. the orthonormal Planck CMB basis. In the remainder of this section, we omit the tilde $``\Tilde{\;\;}"$ and the superscript $``X"$ for simplicity. 

Previously, we have defined the inner product between two mode functions, and the inner product between two bispectra can be defined in the same way as 
\begin{equation}
S_{\mathrm{th}}\cdot S_{\mathrm{data}} \equiv \langle S_{\mathrm{th}}, S_{\mathrm{data}} \rangle_l = \sum_{m,n}\alpha_m\beta_n \langle R_m, R_n \rangle_l = \sum_{n}\alpha_n \beta_n \equiv \alpha_n \beta_n.
\label{E3.4}
\end{equation}
In the third equality, we use the fact $\langle R_m, R_n \rangle_l = \delta_{mn}$, and in the fourth equality, we use the index notation to simplify the expression. 
The shape correlator $\mathcal{S}$ and amplitude correlator $\mathcal{A}$ in \cite{Clarke:2020znk} can therefore be expressed in the \texttt{Modal} language as
\begin{align}
    \mathcal{S}(S_1, S_2) &\equiv \frac{S_1\cdot S_2}{\sqrt{(S_1\cdot S_1)(S_2\cdot S_2)}} = \frac{\alpha_n \beta_n}{\sqrt{(\alpha_m \alpha_m)(\beta_l \beta_l)}}\label{E3.5},
    \\[0.5em]    \mathcal{A}(S_1, S_2) & \equiv \sqrt{\frac{S_1\cdot S_1}{S_2 \cdot S_2}} = \sqrt{\frac{\alpha_m \alpha_m}{\beta_n \beta_n}}.
    \label{E3.6}
\end{align}
We can directly estimate $c_{\mathrm{NL}}$ with equation (\ref{E3.1})
\begin{equation}
    c_{\mathrm{NL}} = \frac{\alpha_n \beta_n}{\alpha_m \alpha_m} = \frac{\mathcal{S}(S_{\mathrm{th}}, S_{\mathrm{data}})}{\mathcal{A}(S_{\mathrm{th}}, S_{\mathrm{data}})},
    \label{E3.7}
\end{equation}
where in the second equality, we use equation (\ref{E3.5}) and equation (\ref{E3.6}).

To estimate the $f_{\mathrm{NL}}$, we firstly define the normalised template as
\begin{equation}
    S_1^{\mathrm{norm}} \equiv \frac{1}{N}S_1 = \sum_n\frac{\alpha_n}{N}R_n \equiv \sum_n\alpha_n^{\mathrm{norm}}R_n    
    \label{E3.8}
\end{equation}

where $N$ is the normalisation factor, with the canonical choice $N = S_{\mathrm{th}}(k_*, k_*, k_*)$. Then $f_{\mathrm{NL}}$ is directly related to $c_{\mathrm{NL}}$ by
\begin{equation}
    f_{\mathrm{NL}} = \frac{\alpha_n^{\mathrm{norm}}\beta_n}{\alpha_m^{\mathrm{norm}}\alpha_m^{\mathrm{norm}}} =\frac{\mathcal{S}(S_1^{\mathrm{norm}}, S_2)}{\mathcal{A}(S_1^{\mathrm{norm}}, S_2)}=Nc_{\mathrm{NL}}.  
    \label{E3.9}
\end{equation}
Noticing that the shape correlator $\mathcal{S}$ is invariant under any overall scaling of the bispectrum, there is $\mathcal{S}(S_1, S_2) = \mathcal{S}(S_1^{\mathrm{norm}}, S_2)$. The difference factor $``N"$ between $f_{\mathrm{NL}}$ and $c_{\mathrm{NL}}$, which equals the normalisation factor, therefore, is completely a consequency of the amplitude correlator.

For template-based analysis, the value of $f_{\mathrm{NL}}$ sensitively depends on the choice of normalisation convention, which determines the amplitude correlator that appears in the denominator of equation (\ref{E3.9}). Instead, the quantity in which people are interested is the SNR: $|f_{\mathrm{NL}}/\sigma_{f}|$. As mentioned above, $\sigma_{f}$ is the standard deviation of 
\begin{equation}
    f_{\mathrm{NL}}^{\mathrm{sim}} \equiv \frac{\alpha_n^{\mathrm{norm}} \beta_n^{s}}{\alpha_m^{\mathrm{norm}}\alpha_m^{\mathrm{norm}}}
    \label{E3.10}
\end{equation}
for 160 CMB maps from gaussian simulations, where $\beta_n^{s}$ are \texttt{Modal} coefficients of the simulation bispectra, with the mean value $\langle\beta_n^s \rangle \approx 0$ for each mode. In the \texttt{Modal} language, 
\begin{equation}
    \sigma_f \equiv \sqrt{\left\langle \left( \frac{\alpha_n^{\mathrm{norm}}(\beta_n^s-\langle\beta_n^s \rangle)}{\alpha_m^{\mathrm{norm}} \alpha_m^{\mathrm{norm}}}\right)^2 \right\rangle} \approx \sqrt{\left\langle \left( \frac{\alpha_n^{\mathrm{norm}}\beta_n^s}{\alpha_m^{\mathrm{norm}} \alpha_m^{\mathrm{norm}}}\right)^2 \right\rangle}= \frac{\sqrt{\langle(\alpha_n^{\mathrm{norm}}\beta_n^s)^2\rangle}}{\alpha_m^{\mathrm{norm}}\alpha_m^{\mathrm{norm}}} = N\frac{\sqrt{\langle(\alpha_n\beta_n^s)^2\rangle}}{\alpha_m\alpha_m},
    \label{E3.11}
\end{equation}
and the significance is given by
\begin{equation}
    \left|\frac{f_{\mathrm{NL}}}{\sigma_f}\right|= \frac{|\alpha_n^{\mathrm{norm}} \beta_n|}{\sqrt{\langle (\alpha_m^{\mathrm{norm}}\beta_m^s)^2\rangle}} = \frac{|\alpha_n \beta_n|}{\sqrt{\langle (\alpha_m\beta_m^s)^2\rangle}}.
    \label{E3.12}
\end{equation}
Explicitly, the expression of $|f_{\mathrm{NL}}/\sigma_f|$ is invariant under any uniform scaling of $\alpha_n$, therefore, the significance is independent of the amplitude correlator. In principle, searching for large significance is equivalent to looking for the large shape correlator between the predicted bispectrum and the data (and small correlation with Gaussian maps to distinguish the PNG from null hypothesis), regardless of the amplitude.

Now we look at the template-free analysis. Similarly, the error of $c_{\mathrm{NL}}$ is defined as
\begin{equation}
    \sigma_c \equiv \sqrt{\left\langle \left( \frac{\alpha_n(\beta_n^s-\langle\beta_n^s \rangle)}{\alpha_m \alpha_m}\right)^2 \right\rangle} \approx \sqrt{\left\langle \left( \frac{\alpha_n\beta_n^s}{\alpha_m \alpha_m}\right)^2\right\rangle}=\frac{\sqrt{\langle(\alpha_n\beta_n^s)^2\rangle}}{\alpha_m\alpha_m},
    \label{E3.13}
\end{equation}
and we have the significance
\begin{equation}
    \left|\frac{c_{\mathrm{NL}}}{\sigma_c}\right| = \frac{|\alpha_n \beta_n|}{\sqrt{\langle (\alpha_m\beta_m^s)^2\rangle}}.
    \label{E3.14}
\end{equation}
Unlike template-based analyses, the template-free approach based on the $c_{\mathrm{NL}}$ parameter can systematically incorporate information about the \emph{amplitude} of the theoretical predictions. Since no normalisation is required when estimating $c_{\mathrm{NL}}$, an ideal theoretical bispectrum would satisfy both $\mathcal{S}(S_1, S_2)\rightarrow 1$ and $\mathcal{A}(S_1, S_2)\rightarrow 1$, yielding $c_{\mathrm{NL}} \approx 1$. Thus, in addition to examining the signal-to-noise ratio (SNR), which constrains only the \emph{shape}, one may impose a selection condition directly on the value of $c_{\mathrm{NL}}$ itself, thereby constraining the \emph{amplitude} of the predicted bispectrum, as captured by our constraining criterion in equation~(\ref{E3.2}).

We illustrate this with a simple example. Consider a scenario predicting
$c_{\mathrm{NL}} = 1.2 \pm 0.3$, 
which satisfies both the constraining and discovery criteria. Now imagine decreasing its amplitude uniformly by a factor of ten while keeping the shape unchanged; equivalently, this corresponds to dividing each modal coefficient $\alpha_n$ by a factor of ten. Owing to the linearity of the \texttt{Modal} estimator, the resulting estimate becomes
\[
c_{\mathrm{NL}} = 12 \pm 3.
\]
This clearly satisfies the discovery criterion---indicating a good shape with a nominal significance of $4\sigma$---but it fails the constraining criterion owing to the bispectrum amplitude being too small. Conversely, if we increase the amplitude by a factor of ten, we obtain
\[
c_{\mathrm{NL}} = 0.12 \pm 0.03,
\]
again implying a good shape but now with an amplitude that is excessively large.

In principle, a scenario can be regarded as a viable candidate only if it satisfies both the constraining and discovery criteria. However, since current observations show no evidence for primordial non-Gaussianity (PNG), we do not expect to obtain a sufficiently large SNR to claim a direct detection at the present stage. Although identifying such models remains our long-term scientific goal, in the test analysis that follows we will not impose the discovery criterion. Instead, we focus solely on the constraining criterion and apply it to the simple IR~DBI model, in order to demonstrate how the pipeline operates and to emphasise that incorporating amplitude information can lead to improved constraints on the underlying theories.

\subsection{Normalise to Local}
In template-based analysis, theoretical templates are normalised by requiring $S^{\mathrm{norm}}(k,k,k)=1$. However, this convention ignores the behaviour away from the equilateral value $k = k_1=k_2=k_3$ and may lead to large disparities between the $f_{\mathrm{NL}}$ constraints for different models. To fix this issue, people proposed the parameter $F_{\mathrm{NL}}$ \cite{Fergusson:2010dm}, which measures the integrated CMB bispectrum signal relative to that of the canonical local model \eqref{2E10} with $f_{\mathrm{NL}}=1$. In \texttt{Modal} language, the $F_{\mathrm{NL}}$ is given by
\begin{equation}
    F_{\mathrm{NL}} = \frac{\alpha_n^{\mathrm{norm}}\beta_n}{\sqrt{(\alpha_m^{\mathrm{norm}}\alpha_m^{\mathrm{norm}})(\alpha_l^{\mathrm{loc}}\alpha_l^{\mathrm{loc}})}},
    \label{5.4E1}
\end{equation}
where $\alpha^{\mathrm{norm}}$ denotes the coefficients of the normalised theoretical template, and $\alpha^{\mathrm{loc}}$ denotes the coefficients of the canonical local template. Both coefficients are in terms of late-time orthonormal basis. Unlike $f_{\mathrm{NL}}$, $F_{\mathrm{NL}}$ is insensitive to the method of normalisation (while it depends on the cut-off scale, or the volume of the tetrapyd domain). Therefore, this measurement accommondate a broader range of models, including those with non-trivial scale-dependence. For template-free analysis, numerical calculation gives the coefficient $\alpha_n$ without any normalisation. $F_{\mathrm{NL}}$ is related to $c_{\mathrm{NL}}$ by
\begin{equation}
    F_{\mathrm{NL}} = \frac{\alpha_n\beta_n}{\sqrt{(\alpha_m\alpha_m)(\alpha_l^{\mathrm{loc}}\alpha_l^{\mathrm{loc}})}}=c_{\mathrm{NL}}\sqrt{\frac{\alpha_n\alpha_n}{\alpha_m^{\mathrm{loc}}\alpha_m^{\mathrm{loc}}}}.
\end{equation}
Using this parameter means that all numerical shapes are normalised to match the constraint of the local shape, which does not fully utilise the numerically computed amplitude of the correlation. However, even within template-free analyses, moderately reducing the emphasis on the amplitude and instead focusing on higher precision shapes/templates from numerical computation may provide new insight which potentially guides the future direction to achieve a significant discovery.

\section{Constraining Inflationary Models: Taking the IR DBI Inflation as An Example}
\label{section:example}
In this section, we perform a proof-of-concept analysis to constrain the inflationary parameter space. Our pipeline is divided into two parts: power spectrum analysis and bispectrum analysis. In the power spectrum analysis, for each scenario within (a range of) parameter space, the equation of motion of curvature perturbation is solved fully numerically with an Lsoda solver to get the predicted scalar power spectrum. Then the amplitude $A_s$ and the spectral index $n_s$ of the scalar power spectrum at the pivot scale $k_* = 0.05$ are compared with the values in the Planck 2018 results for T+E, which are $A_s = 2.1\times10^{-9}$ and $n_s = 0.9649\pm0.0042$ \cite{Planck:2018jri}. Scenarios with both amplitude and spectral index consistent with the Planck results (within $1\sigma$) are then proceeded to the bispectrum analysis, as discussed in section \ref{sect:c_nl}. The joint-analysis of the power spectrum and the bispectrum is expected to rule out most scenarios, and the ensemble of surviving scenarios can provide constraints on the inflationary parameter space. We validate focus on the simplest model of IR DBI inflation introduced in section \ref{section: model}.
\subsection{Parameter Space}
The model is presented by equation (\ref{4E1}), where the Lagrangian is uniquely determined by three model parameters: $\lambda$, $V_0$ and $\beta$. To set the intial condition, two other (model-independent) parameters are considered, which are $\phi_{i}$, the initial value of the inflaton field and $\Delta N_s$, the number of e-folds to the end of inflation since the mode with pivot scale $k_* = 0.05$ crosses the Hubble radius, where we assume the inflation ends at 70 e-folds. Once $\phi_i$ is fixed, all other dynamical quantities, such as $\phi'$ and $H$ can be evaluated at the initial time using slow-roll Hubble constraints. $\Delta N_*$ is used to normalise the scale factor, in other words, to fix the correspondence between the wavenumber $k$ and the predicted value of the spectrum \cite{Kinney:2005in}. We assume a canonical post-inflation evolution of the universe.

Therefore, constraining the inflationary model requires computing the power spectra and bispectra for scenarios spanning a five-dimensional parameter space. Although \texttt{Primodal} can evaluate an individual bispectrum efficiently---typically within three minutes for most scenarios, each sampled on a $400\times400\times400$ grid---scanning a large five-dimensional parameter space remains computationally demanding and impractical at this stage. For the more modest goal of validating the pipeline, we therefore restrict our analysis to a relatively small region surrounding a carefully chosen benchmark scenario:

\begin{align}
\lambda &= 2.00475 \times 10^{15}
\label{6E1.1}
\\ V_0 &= 5.2\times10^{-12}~\mathrm{M_{pl}}^4
\label{6E1.2}
\\ \beta &= 0.29
\label{6E1.3}
\\ \phi_i &= 0.46042~\mathrm{M_{pl}}
\label{6E1.4}
\\ \Delta N_* &= 55.8
\label{6E1.5}
\end{align}
which predicts a scalar power spectrum with $As = 2.099\times10^{-9}$ and $n_s = 0.9647$, in agreement with Planck's 2018 result. 

In the string-inspired model, IR DBI inflation ends at $\phi_e \approx H\sqrt{\lambda}$ \cite{Bean:2007eh}, and in the scenario mentioned above, this relation predicts the stringy end of inflation at $\phi_e \approx 58$, which is about 50 times greater than the value of the inflaton field at 70 e-folds, when we force the inflation to end. This means that our choice of $N_{\mathrm{tot}} = 70$ makes inflation end deeply inside the IR DBI regime in this scenario and can, therefore, avoid the extra difficulty introduced by the transition between the DBI phase (kinetic dominate) and the non-relativistic phase (potential dominate) \cite{Bean:2007eh} \cite{Miranda:2012rm}. 

The region for our parameter scan is determined as
\begin{align}
\lambda &\in [1.9 \times 10^{15}, 2.1 \times 10^{15}];
\label{6E2.1}
\\ V_0 &\in [5.0\times10^{-12}, 5.4\times10^{-12}]~\mathrm{M_{pl}}^4;
\label{6E2.2}
\\ \beta &\in [0.16, 0.6];
\label{6E2.3}
\\ \phi_i &\in [0.455, 0.465]~\mathrm{M_{pl}};
\label{6E2.4}
\\ \Delta N_* &\in [55, 56.5].
\label{6E2.5}
\end{align}
The boundary values of all parameters except $\beta$ correspond to roughly 10\% variations in $A_s$ and $n_s$ compared to the central value, and varying $\beta$ approximately leaves the power spectrum unchanged \cite{Clarke:2020znk}. In contrast, from the analytic expression of the DBI bispectrum in equation (\ref{4E3}) and the fact that $c_s\approx (\beta N/3)^{-1}$, people notice that the amplitude of the bispectrum is mainly controlled by the value of $\beta$. As a result, we expect that the main constraint on $\beta$ comes from the bispectrum, and the main constraints on the other parameters come from the power spectrum. 

One disappointing fact is that the region of parameter space that we will scan over is very narrow, so it is unlikely that we can get "global" constraints from it. Limited by current computational power, we are unable to scan a reasonably larger parameter space which matches the theoretical bounds of these parameters. We perform this 5-dimensional parameter scan to demonstrate how our pipeline works and prove its potential to be used in future practical analysis, and it can also give us straightforward visualisation of the configuration of the parameter space constrained by the Planck power spectrum and bispectrum, providing guidance on further numerical investigation. In section \ref{go outside}, we use the monotonic dependence between the power spectrum/bispectrum and each parameter in this simple inflation model to go beyond this narrow region and try to find physically meaningful constraints.

\subsection{The Power Spectrum Analysis}
660000 scenarios are inputted into the pipeline, separated uniformly. For each scenario, the equation of motion of curvature perturbations is solved numerically with a Lsoda solver, and the tree-level scalar power spectrum is calculated with \cite{Planck:2013jfk}
\begin{equation}
    \mathcal{P}_{\zeta} = \frac{k^3|\zeta_{k}|^2}{2\pi^2}
    \label{6E3}
\end{equation}
at $\tau \approx 0$. Afterwards, the amplitude $A_s$ and the spectra index $n_s$ are evaluated by fitting \cite{Planck:2018jri}
\begin{equation}
    \log\mathcal{P}_{\zeta} = \log A_s + (n_s-1)\times(\log k-\log k_{*})
    \label{6E4}
\end{equation}
to the numerical result and are compared with the literature. 

Scenarios that predict amplitudes and spectra indices satisfying
\begin{align}
A_s &\in [2.070\times10^{-9}, 2.128\times10^{-9}]
\\n_s &\in [0.9607, 0.9691]
\label{6E5}
\end{align}
which corresponds to the $1\sigma$ region of Planck's observational result \cite{Planck:2018jri}, are selected as "good scenarios" and are forwarded to the bispectrum analysis. The rest of the scenarios are ruled out because they do not predict correct power spectra. 

Unlike the bispectrum, the analysis techniques for the power spectrum are very sophisticated, enabling people to project the primordial power spectrum to late time and obtain constraints from the full shape of the power spectrum. However, in this work, for the purpose of validating and emphasising the bispectrum part of the conceptual pipeline, we only consider the constraints of the amplitude and spectra index of the primordial power spectrum. In practical investigations, people should use power spectrum codes like $\texttt{CAMB}$ \cite{Challinor} for a systematic analysis.

\subsection{The Bispectrum Analysis}
For each scenario that survives in the analysis of the power spectrum, its tree-level bispectrum is calculated by \texttt{Primodal}. In addition, the value of $c_{\mathrm{NL}}$, as well as its standard deviation $\sigma$ is evaluated with equation (\ref{E53}). As explained in section \ref{sect:c_nl}, scenarios are ruled out by ruling out $c_{\mathrm{NL}} = 1$ from their predicted $2\sigma$ region. 

We can directly read out the constraints on inflationary parameter space from the remaining scenarios, which haven't been ruled out by the scalar power spectrum and bispectrum information from observational data. For all remaining scenarios, an investigation of their fitting significance, $|c_{\mathrm{NL}}/\sigma|$, is carried out to obtain further implications on the constraints of parameter space from the shape of the bispectrum.

\subsection{Results of the Parameter Scan}
After power spectrum analysis, 566486 of the 660000 scenarios are ruled out by $A_s$ and $n_s$. For the 93514 scenarios predicting power spectra consistent with observation, 48263 are further ruled out by $c_{\mathrm{NL}}$ from bispectrum analysis. 

\begin{figure}
    \centering
    \begin{subfigure}{0.8\textwidth}
        \centering
        \includegraphics[width=\columnwidth]{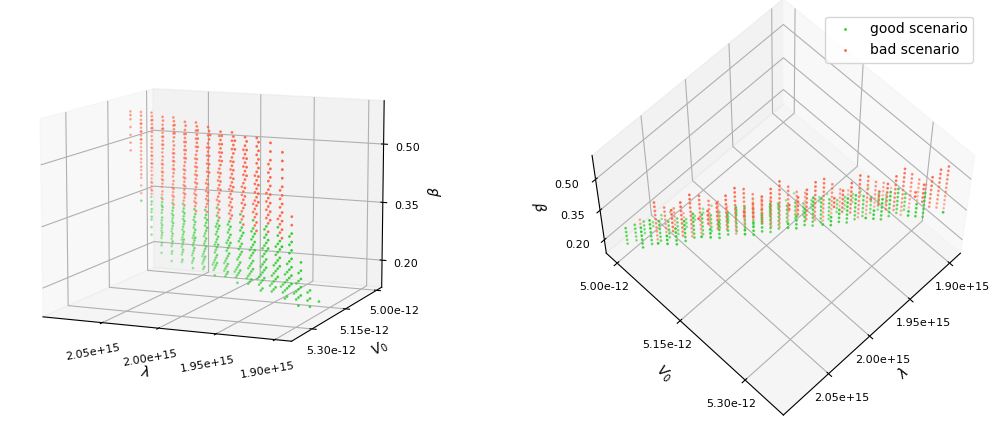}
        \caption{$\phi_i = 0.457$, $\Delta N_* = 55.4$}
        \label{F1.1}
    \end{subfigure}
    \newline
    \centering   
    \begin{subfigure}{0.8\textwidth}
        \centering
        \includegraphics[width=\columnwidth]{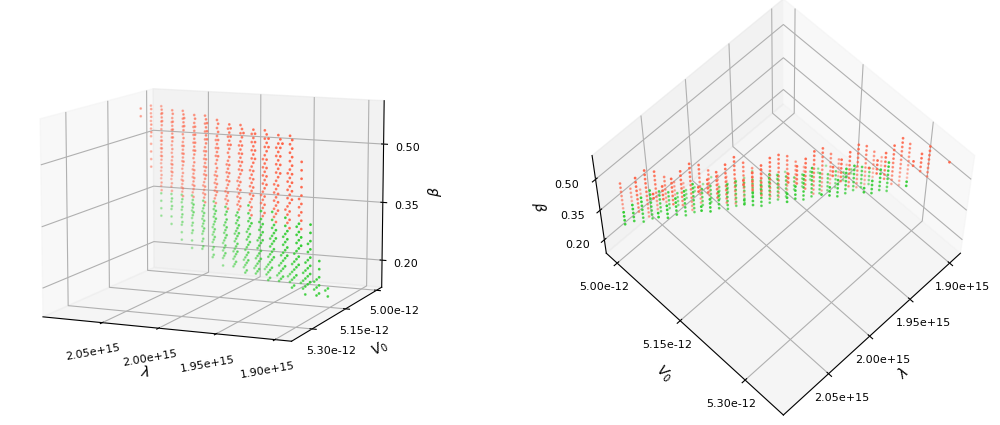}
        \caption{$\phi_i = 0.457$, $\Delta N_* = 56.2$}
        \label{F1.2}
    \end{subfigure}
    \newline
    \centering
    \begin{subfigure}{0.8\textwidth}
        \centering
        \includegraphics[width=\columnwidth]{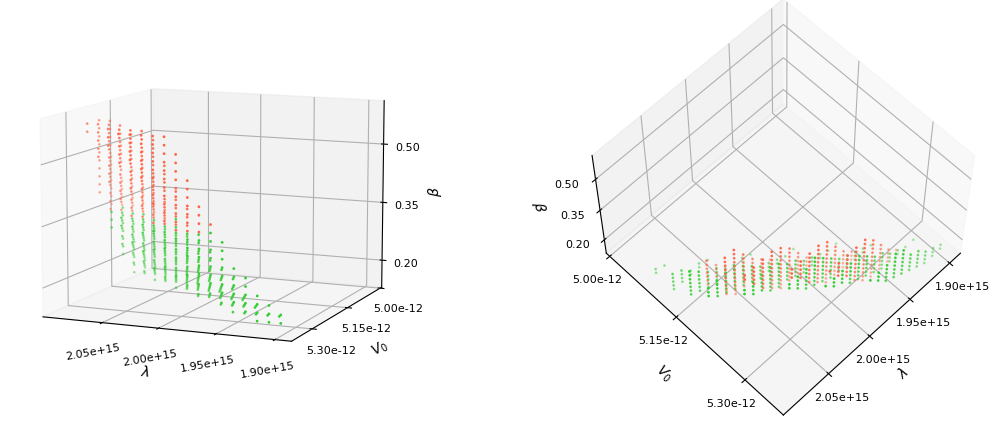}
        \caption{$\phi_i = 0.463$, $\Delta N_* = 55.4$}
        \label{F1.3}
    \end{subfigure}
    \newline
    \caption{Scenarios in the sub-parameter space of $\lambda$, $V_0$ and $\beta$, at (a) $\phi_i = 0.457$, $\Delta N_* = 55.4$; (b) $\phi_i = 0.457$, $\Delta N_* = 56.2$; (c) $\phi_i = 0.457$, $\Delta N_* = 55.4$. Scatters (red and green) correspond to scenarios with $A_s$ and $n_s$ located inside the $1 \sigma$ region of Planck 2018 results for the scalar power spectrum. Red dots label scenarios ruled out by bispectrum analysis (with $c_{\mathrm{NL}} = 1$ being outside their $2 \sigma$ region). Green dots label scenarios which cannot be ruled out by $2\sigma$.}
    \label{F5.1}
\end{figure}

\begin{figure}
     \centering
     \includegraphics[width=\columnwidth]{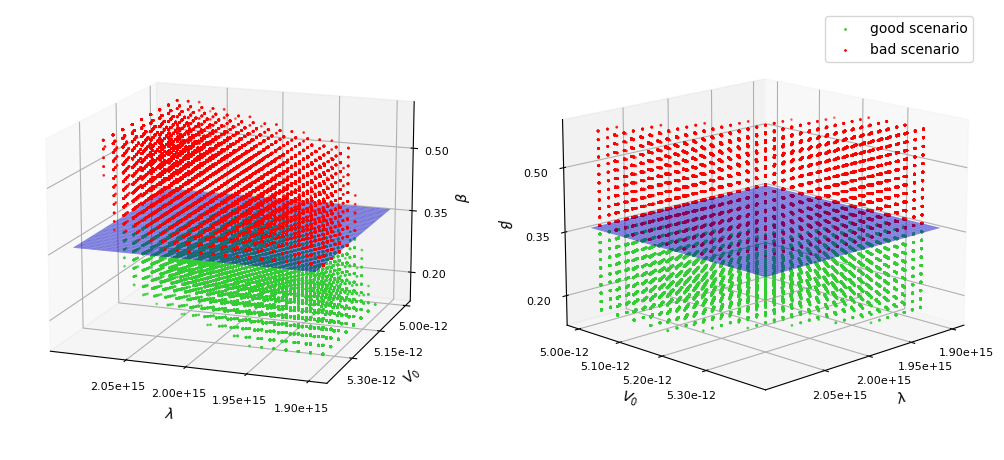}
     \caption{All secnarios in the sub-parameter space of $\lambda$, $V_0$ and $\beta$ favoured by power spectrum, marginalizing over $\phi_i$ from 0.455 to 0.465 and $\Delta N_*$ from 55 to 56.5. Green scatters label scenarios that survive after the bispectrum analysis, and red scatters label those inconsistent with the Planck bispectrum by $2\sigma$ and are therefore ruled out. The Blue plane corresponds to $\beta = 0.36$, which is interpreted as the constraint on $\beta$.}
     \label{F5.2}
\end{figure}

To visualise the results of the power spectrum scan and the bispectrum scan, the reduced 3-d parameter space of $\lambda$, $V_0$, $\beta$, with fixed $\phi_i$s and $\Delta N_*$s are plotted in Figure \ref{F5.1}. We notice that all scenarios allowed by the scalar power spectrum, with fixed $\Delta N_*$ and $\phi_i$, are constrained in a narrow region on the constant-$\beta$ plane (and constant $\Delta N_*$, $\phi_i$ too), which implies the strong constraining power of the Planck power spectrum in the warp space geometry (encoded in $\lambda$) and the energy scale of inflation (encoded in $V_0$). In contrast, the constraint on $\beta$, which directly relates to the sound speed $c_s$ and the non-Gaussianity amplitude $f_{\mathrm{NL}}$, comes mainly from bispectra. Figure \ref{F5.1} shows clear boundaries between scenarios ruled out and not ruled out by bispectra. This boundary is located around $\beta = 0.36$ and is almost independent of the variation of other parameters. To see that clearly, we plot all scenarios favoured by the Planck power spectrum in Figure \ref{F5.2}, marginalising $\phi_i$ and $\Delta N_*$. Therefore, we take this flat boundary as the constraint on $\beta$, which is $\beta \leq 0.36$. 

\begin{figure}
     \centering
     \includegraphics[width=\columnwidth]{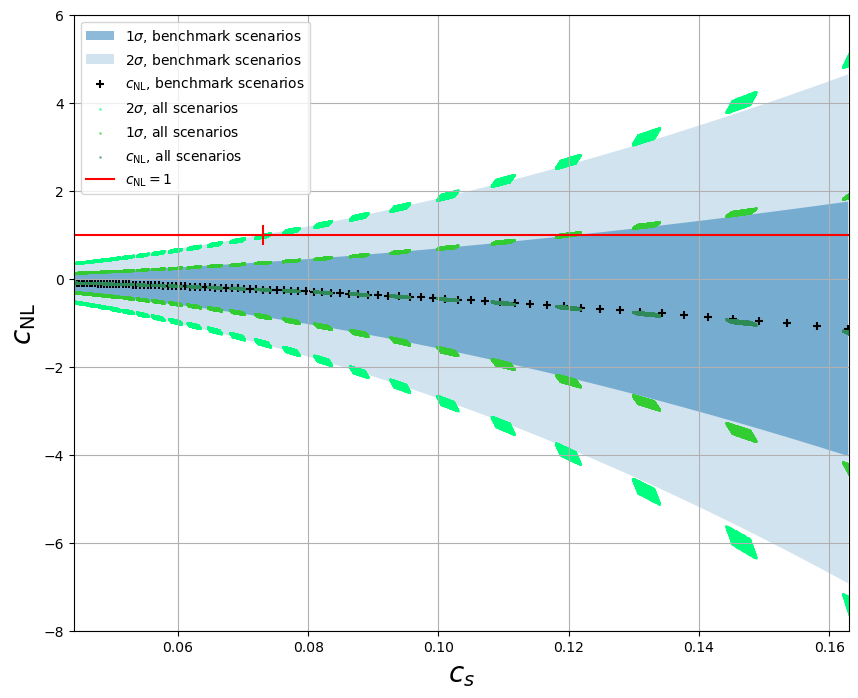}
     \caption{The quantity $c_{\mathrm{NL}}$ is plotted as a function of $c_s$. The ``+'' symbols mark the values of $c_{\mathrm{NL}}$ for the benchmark scenarios, in which only $\beta$ varies from 0.16 to 0.6, while all other parameters are fixed according to Eqs.~(\ref{6E1.1})--(\ref{6E1.5}). (Note that some of these benchmark scenarios are excluded by the power-spectrum constraints.) The corresponding $1\sigma$ and $2\sigma$ confidence regions are shown as the dark--blue and light--blue shaded areas, respectively. The green scatter points indicate the values of $c_{\mathrm{NL}}$, together with $c_{\mathrm{NL}}\pm\sigma$ and $c_{\mathrm{NL}}\pm2\sigma$, obtained from the full five--dimensional parameter scan after imposing the power-spectrum constraints. The red horizontal line corresponds to $c_{\mathrm{NL}} = 1$. It intersects the boundary of the $2\sigma$ region of the power-spectrum--allowed scenarios at $c_s = 0.073$, which is therefore interpreted as the lower bound on the sound speed.}
     \label{F5.3}
\end{figure}

\begin{figure}
     \centering
     \includegraphics[width=\columnwidth]{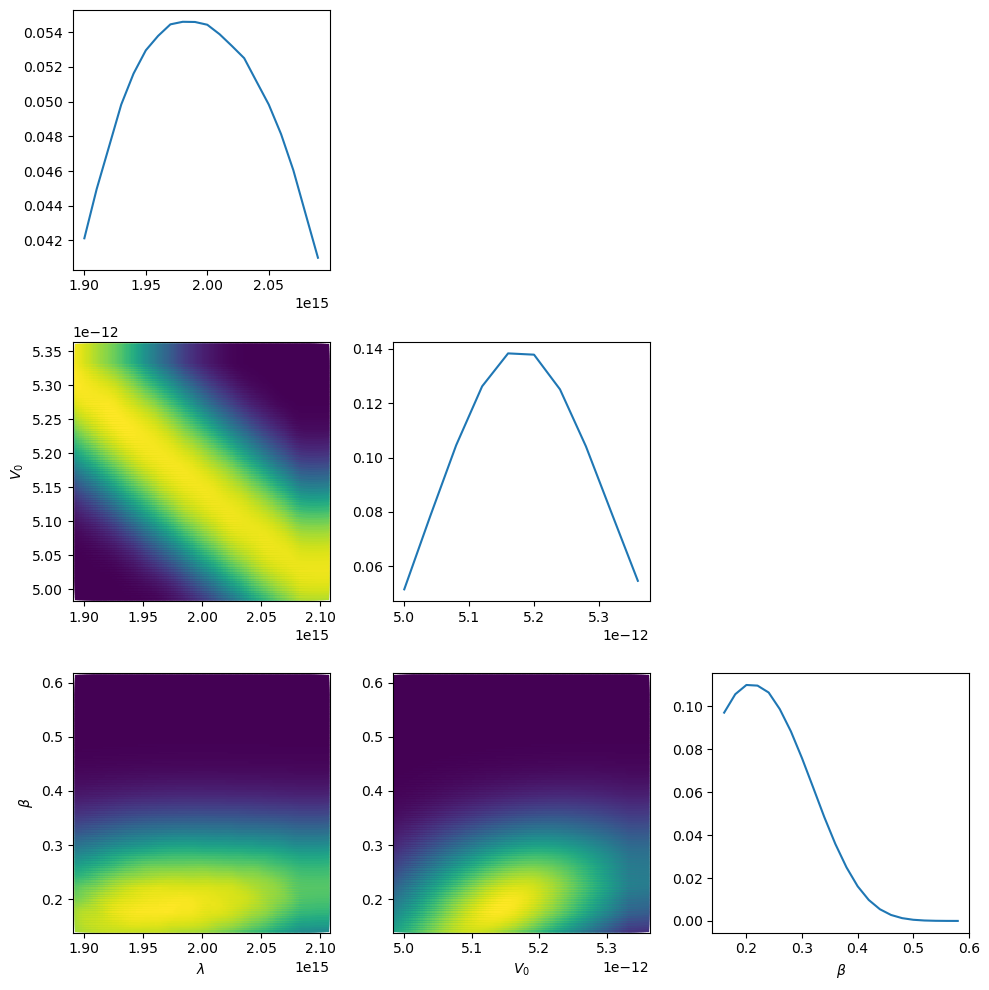}
     \caption{Triangle plot for the marginalised likelihood distributions for three model parameters of the IR-DBI inflation. In 2-dimensioanl plots, bright colours region represent scenarios with large likelihood.} 
     \label{F5.6}
\end{figure}

For all scenarios allowed by the power spectrum, we solve their equations of motion of the background inflaton field numerically, and these numerical solutions can tell us the values of sound speed $c_s$ when the pivot mode crosses the horizon. Figure \ref{F5.3} shows $c_{\mathrm{NL}}$ vs $c_s$, from which we get the constraint on sound speed by finding the intersection between $c_{\mathrm{NL}} = 1$ and the boundary of $2 \sigma$ region.
This approach determines the constraint on sound speed as $c_s \geq 0.073$.
The likelihood for scenarios satisfying the bispectrum constraints are computed with
\begin{equation}
    \mathcal{L}\equiv\frac{1}{\sqrt{2\pi\sigma}}e^{-\frac{(c_{\mathrm{NL}}-1)^2}{2\sigma^2}},
    \label{6.4E1}
\end{equation}
(while the likelihoods for the scenarios ruled out by power spectrum are set to be 0). The marginalised likelihood distributions are plotted in the triangle form in Figure \ref{F5.6}.
In selected region, $\lambda$ and $V_0$ have Gaussian-distribution while $\beta$ shows a deviation, which agrees with the constraining cutoff in the three-dimensional plots in Figure (\ref{F5.1}) and Figure (\ref{F5.2}).

\begin{figure}
     \centering
     \includegraphics[width=\columnwidth]{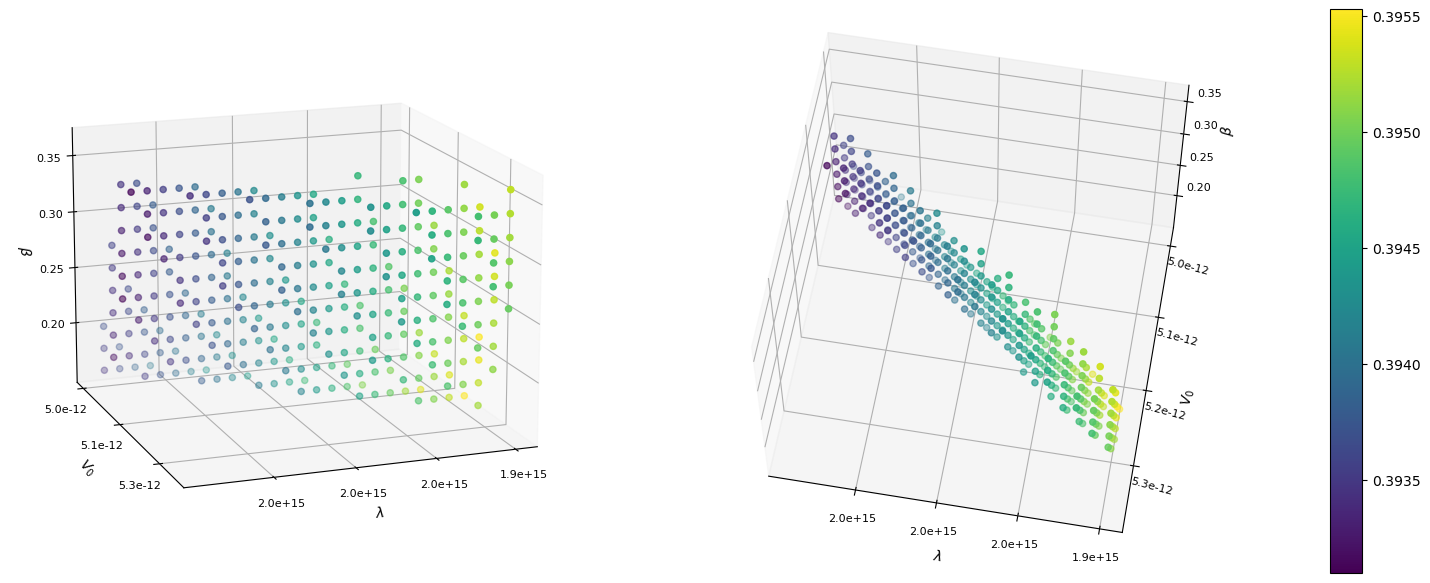}
     \caption{Significance--$|c_{\mathrm{NL}}/\sigma|$ for scenarios \textbf{not} excluded by $A_s$, $n_s$ of the Planck power spectra and $c_{\mathrm{NL}}$ from the Planck bispectrum. Two initial condition parameters are fixed as $\phi_i = 0.460$ and $\Delta N_* = 55.8$. The colour map labels the significance value, varying from 0.3931 to 0.3955.}
     \label{F5.4}
\end{figure}

As discussed in section \ref{sect:c_nl}, the significance, which is given by $|\frac{c_{\mathrm{NL}}}{\sigma}|$, generally tells us how well the \textit{shape} of the predicted bispectrum fits the data, and how unlikely that the predicted primordial bispectrum to be observed in a "Gaussian" universe. Therefore, significance examination plays an important role in further constraining the parameter space, especially to exclude scenarios predicting a large $\sigma$ due to the bad shape, but cannot be ruled out by $|c_{\mathrm{NL}}-1|<2\sigma$. For scenarios passing both the power spectrum and the bispectrum analysis, their significance is plotted in Figure \ref{F5.4}. One can easily find that significance only varies in a small range over these scenarios, from 0.3931 to 0.3955, which means that none of these scenarios satisfies the discovery criterion in equation (\ref{E3.3}) and does not show evidence of non-Gaussianities from DBI inflation. However, it is still worth commenting on the numerical result of the significance analysis because of its implication on how different parameters affect the bispectrum prediction in DBI inflation. From Figure \ref{F5.4}, we notice that the value of significance does not show obvious dependence on $\beta$. In contrast, a weak but clear dependence on $\lambda$ and $V_0$ can be observed -- the significance increases as $\lambda$ decreases and $V_0$ increases. We do not try to improve the constraints from the significance considering its slow varying behaviour in the parameter space we scanned over, but these plotting results still validate our expectations: Unlike the amplitude, the shape of the DBI bispectrum, or more concretely, the scale dependence of non-Gaussianity, is dominantly controlled by $\lambda$ and $V_0$.  Moreover, the variation tendency of significance implies the location of the best-fitted scenario, so further investigation can be carried out along the trajectory on the $\lambda$ and $V_0$ plane to attempt to find the best-fitted results, and this is actually the motivation of developing the Markov-Chain-Monte-Carlo pipeline for bispectrum analysis. However, it is unlikely to find a scenario with a significant result from this model without introducing any feature, and doing analysis in a greater parameter space, which is time-consuming and beyond the validation purpose of this paper.

\subsection{Go Outside the Narrow Region}
\label{go outside}
\begin{figure}
    \centering
    \begin{subfigure}{0.7\textwidth}
        \centering
        \includegraphics[width=\columnwidth]{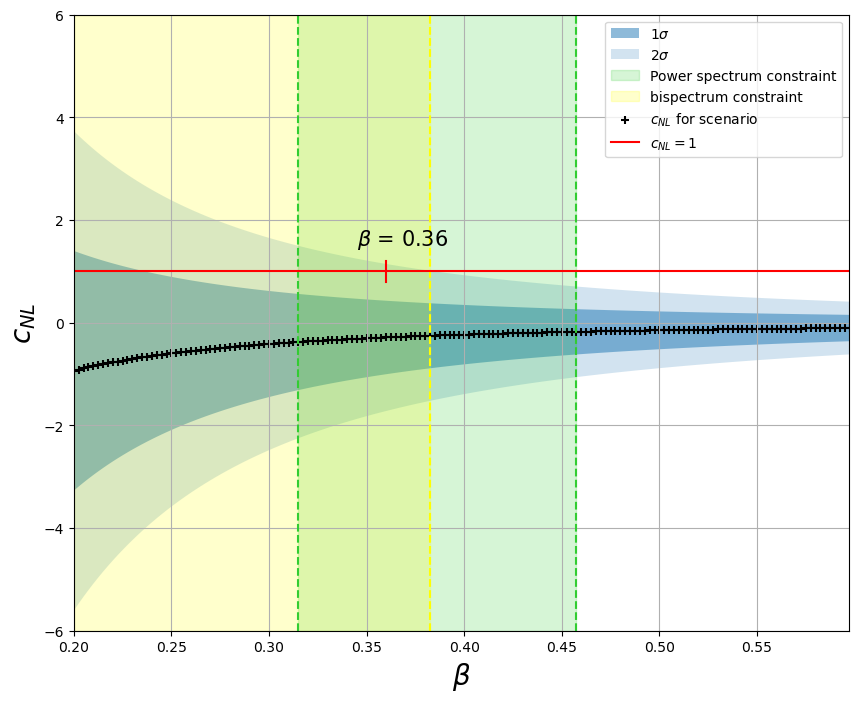}
        \caption{$\lambda = 1.4 \times 10^{15}$, $V_0 = 8 \times 10^{-12}$, $\phi_i = 0.49$, $\Delta N_* = 50$}
        \label{F5.5.1}
    \end{subfigure}
    \newline
    \centering
    \begin{subfigure}{0.7\textwidth}
        \centering
        \includegraphics[width=\columnwidth]{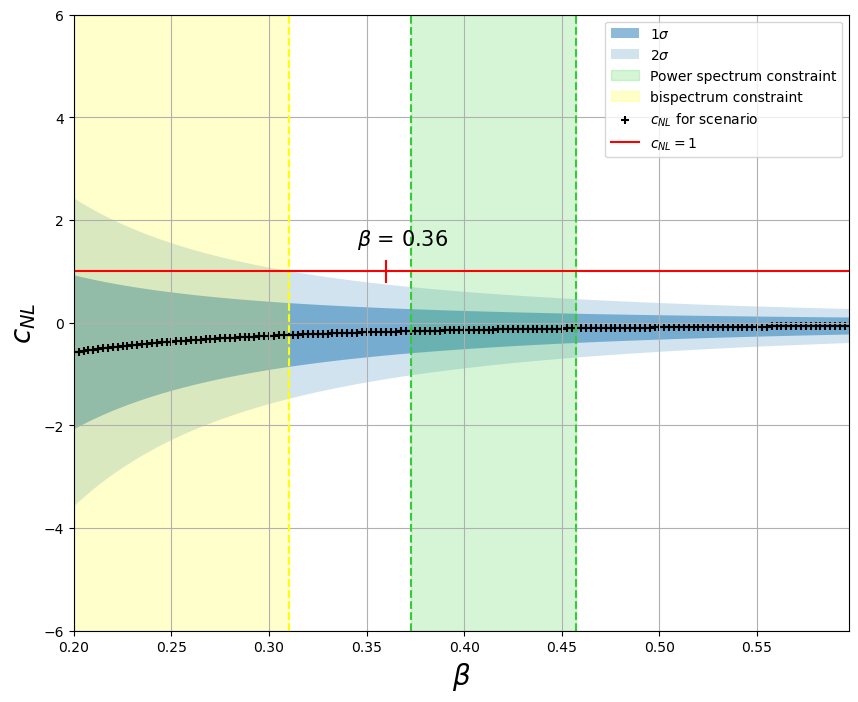}
        \caption{$\lambda = 3.5 \times 10^{15}$, $V_0 = 3.5 \times 10^{-12}$, $\phi_i = 0.45$, $\Delta N_* = 59$}
        \label{F5.5.2}
    \end{subfigure}
    \newline
    \caption{$C_{NL}$ vs $\beta$. "+" symbols label scenarios with $\beta$ varying from 0.2 to 0.6, and with other parameters fixed at: (a) $\lambda = 1.4 \times 10^{15}$, $V_0 = 8 \times 10^{-12}$, $\phi_i = 0.49$, $\Delta N_* = 50$; (b) $\lambda = 3.5 \times 10^{15}$, $V_0 = 3.5 \times 10^{-12}$, $\phi_i = 0.45$, $\Delta N_* = 59$. Dark and light blue areas represent $1\sigma$ and $2\sigma$ regions for scanned scenarios. Green and yellow areas represent the constraints on $\beta$ from the Planck scalar power spectrum and bispectrum, respectively. In case (a), the Planck power spectrum requires $ 0.32< \beta < 0.46$, and the Planck bispectrum requires $\beta < 0.39$; in case (b), the Planck power spectrum requires $ 0.37 < \beta < 0.46$, and the Planck bispectrum requires $\beta < 0.31$.}
    \label{F5.5}
\end{figure}

Limited by current computational resources, we can only scan over a limited region of parameter space. The full practical analysis requires scanning over a wide range of parameter space with Markov chains Monte Carlo (MCMC), however, in this case, benefiting from the simplicity of the infation model, we could explore a wider region by case-by-case study along with 1-dimensional parameter scans. We apply the following strategy: Firstly, we specify a value of $\lambda$ to be outside the previously scanned region and find one scenario that satisfies the Planck power spectrum in $A_s$ and $n_s$ by hand. This time, $V_0$ can take any value, but $\phi_i$ and $\Delta N_*$ are confined in the ranges of $\phi_i \in [0.45, 0.49]\mathrm{M_{pl}}$ and $\Delta N_* \in [50, 60]$, which are consistent with the theoretical bounds on reheating dynamics\cite{Planck:2018jri}. Once such a scenario is found, we perform a 1-dimensional parameter scan of $\beta$ from 0.16 to 0.6 and keep other parameters fixed, then confront numerically computed bispectra to the data with the \texttt{Primodal}+\texttt{Modal} pipeline to investigate how the boundary on $\beta$ between scenarios favoured and disfavoured by the Planck bispectrum deviate from our previous constraint: $\beta = 0.36$. We start from the upper/lower boundary of $\lambda$ of the previously scanned region, then gradually increase/decrease $\lambda$ to step away and repeat the above procedures for each $\lambda$. We notice that for $\lambda > 3.5 \times 10^{15}$ and for $\lambda < 1.4 \times 10^{15}$, it is unlikely to find any scenario satisfying the Planck power spectrum constraints, so we take $1.4 \times 10^{15}$ and $3.5 \times 10^{15}$ as the lower and upper bounds for $\lambda$ from the Planck power spectrum, respectively. The results of $\beta$-scan with $\lambda = 1.4 \times 10^{15}$ and $\lambda = 3.5 \times 10^{15}$ are demonstrated in Figure \ref{F5.5}. The plots are in the same manner as in Figure \ref{F5.3}, and the only difference is that the horizontal axis expresses the value of $\beta$ this time instead of the sound speed.

As illustrated in Figure \ref{F5.5.2}, for a greater $\lambda$, we find that the boundary value of $\beta$ decreases, which is $\beta = 0.31$ for $\lambda = 3.5 \times 10^{15}$ and continues to decrease with increasing $\lambda$. This means that our previous constraint $\beta \leq 0.36$ is roughly safe for scenarios with greater $\lambda$; more concretely, it is unlikely to find a scenario with $\beta > 0.36$ and $\lambda > 1.9 \times 10^{15}$ preferred by both the Planck power spectrum and the bispectrum. Notice that in the case of Figure \ref{F5.5.2}, there is no overlap between the range of $\beta$ allowed by the Planck power spectrum (marked as green) and the Planck bispectrum (marked as yellow), so all these scenarios must be ruled out by the joint analysis. Inversely, for smaller $\lambda$, the boundary value of $\beta$ increases as $\lambda$ decreases, illustrated in the example shown in Figure \ref{F5.5.1}. This time, the overlap between the green and yellow areas corresponds to scenarios that cannot be ruled out by the joint analysis, and some of them have $0.36 <\beta < 0.39$. Therefore, the constraint on $\beta$ needs to be loosened as scenarios with small $\lambda$ possibly break the previous constraint, $\beta < 0.36$, 

The boundary value on the sound speed $c_s$ is found to be less sensitive to the variation of $\lambda$ and $V_0$ compared to $\beta$. The Planck bispectrum requires $c_s > 0.074$ and $c_s > 0.073$ for scenarios with $\lambda = 1.4 \times 10^{15}$ and $\lambda = 3.5 \times 10^{15}$, respectively, in the above analysis. Therefore, our previous constraint $c_s \geq 0.073$ still holds in this broader region.

In summary, we broaden the investigation to a wider parameter space, which is  
\begin{align}
\lambda &\in [1.4 \times 10^{15}, 3.5 \times 10^{15}];
\label{6E6.1}
\\ V_0 &\in [3.5\times10^{-12}, 8.0\times10^{-12}] M_{pl}^4;
\label{6E6.2}
\\ \beta &\in [0.16, 0.6];
\label{6E6.3}
\\ \phi_i &\in [0.45, 0.49] M_{pl};
\label{6E6.4}
\\ \Delta N_* &\in [50, 60].
\label{6E6.5}
\end{align}
Within this region, our previous constraint on the sound speed $c_s \geq 0.073$ is still valid, but there is evidence to loosen the constraint on $\beta$ to $\beta < 0.39$. We take these as our final results because outside this parameter region, considering the theoretical bounds on the value of the inflaton field and the duration of inflation, our case-by-case analysis shows there is no scenario predicting compatible amplitude and spectral index with the Planck power spectrum, as a consequence of the monotonic parameter dependence of the power spectrum of this model. However, we must acknowledge that the new boundaries of the parameter space are not obtained from a systematic 5-dimensional parameter scan due to the limitation of computational power, so they can only be regarded as approximations. Furthermore, the boundary values of $\lambda$, $V_0$ are sensitively dependent on the theoretical bounds of $\phi_i$ and $\Delta N_*$ under the assumption of instantaneous reheating, which means if some non-canonical physics happening during the reheating process change the time duration and energy scale significantly, the range of $\lambda$ and $V_0$ should also be changed. Fortunately, the bispectrum and the boundary on $\beta$ are insensitive to these changes in this simple and nearly scale-independent DBI model. Therefore, we can still treat our modified constraints on $\beta$ and the DBI sound speed as physically meaningful results, at least on some specific energy scale. In the future, with optimised \texttt{Primodal} code and richer computational resources, people can naturally obtain accurate constraints for general inflation models by exploring a reasonably ample parameter space with machine learning and the MCMC method.

\section{Conclusions and Discussions}
\label{section:conclusion}
In this work, we propose a consistency indicator $c_{\mathrm{NL}}$, which shows how well the numerically predicted bispectrum fits the data. Based on this indicator and the significant enhancement of the calculation efficiency of primordial bispectrum brought by \texttt{Primodal}, we propose a template-free pipeline to constrain the parameter space of inflationary Lagrangians. We demonstrate and validate our conceptual pipeline using the Planck 2018 T+E data to constrain the sound speed and $\beta$ parameter of the IR DBI inflation.

In the pipeline, the tree-level scalar power spectra of all scenarios are calculated by solving the equation of motion of the curvature perturbation $\zeta$ numerically, and scenarios with $A_s$ and $n_s$ incompatible with the Planck 2018 results by $1\sigma$ are ruled out. For scenarios passing the power spectrum analysis, their tree-level bispectra are calculated by \texttt{Primodal} and confronted with data with a \texttt{Modal}-like $c_{\mathrm{NL}}$ estimator. The parameter space is further restricted by ruling out scenarios with $c_{NL} = 1$ lying outside their $2 \sigma$ regions. \texttt{Primodal} can directly output the predicted bispectrum as mode expansion coefficients. With precalculated $\sigma$ and $\Gamma$ matrices, the coefficients of the primordial bispectrum will be efficiently projected to $l$-space and transformed to the Planck CMB basis, then confronted to the data with Equation (\ref{E53}) to estimate $c_{\mathrm{NL}}$. That means once \texttt{Primodal} finishes the bispectrum calculation, the remaining work of the analysis is just a few simple matrix multiplications, which require minimal complications. With the current version of \texttt{Primodal}, it takes about 6 minutes to evaluate the $c_{\mathrm{NL}}$ from the DBI Lagrangian for a single scenario, although this may take longer for some scale-dependent single-field inflation models depending on the dimensions of the parameter space. Therefore, further optimisation of the integration code is still desired for practical purposes.

It is worthwhile to mention that we use different confidence levels in power spectrum analysis and bispectrum analysis ($1\sigma$ for power spectrum and $2\sigma$ for the bispectrum). That is because we want some flexibility to vary the bispectrum while keeping the power spectrum relatively fixed to validate the effectiveness of the bispectrum part of the pipeline. Meanwhile, the sound speed and the $\beta$ parameter are the main targets of the bispectrum analysis, which are only weakly dependent on the constraints of the power spectrum. However, for practical analysis in future, people should use consistent confidence levels in the joint analysis.

By scanning over a region in the parameter space defined by equation (\ref{6E2.1}) to (\ref{6E2.5}), we find a constraint on the sound speed (at horizon crossing of pivot modes with $k_* = 0.05$) of 
\begin{equation}
    c_s \geq 0.073 \qquad (95\%,\quad T+E).
\label{7E1}
\end{equation}
and a constraint on $\beta$ of 
\begin{equation}
    \beta \leq 0.39 \qquad (95\%,\quad T+E).
\label{7E2}
\end{equation}

The idea of our work is inspired by a similar validation work performed by Clarke in his Ph.D. thesis \cite{Clarke:2020znk}. This scans $\beta$ across the range
\begin{equation}
    \beta \in [0.19, 0.58].
\label{7E3}
\end{equation}
while keeping all other parameters fixed. For each scenario in this $\beta$ scan, $c_{\mathrm{NL}}$ is computed (which is called $\Tilde{f}_{\mathrm{NL}}$) with \texttt{Primodal}. Scenarios with $c_{\mathrm{NL}} = 1$ being outside their $2 \sigma$ region are ruled out, with the constraint on the DBI sounds speed of
\begin{equation}
    c_s \geq 0.056 \qquad (95\%,\quad T,\quad Clarke).
\label{7E4}
\end{equation}
from Planck 2018 T data only.

Compared with this previous work, our work includes variations of all parameters. Although we expect that the constraints on sound speed and the amplitudes of non-Gaussianities dominantly controlled by $\beta$, allowing all parameters to vary can help us go beyond the theoretical level to investigate the $\lambda$ and $V_0$ dependence of $c_{s}$ and $c_{NL}$. One key limitation of Clarke's work is that his constraint on sound speed can not be directly inverted to the constraints on parameter space. That is because he does not have a mapping from the scales during inflation and the observable scales today, or more specifically, he does not consider the problem of establishing a correspondence between the value of the spectrum to the observational wave number, as mentioned in \cite{Clarke:2020znk}. Our work partially solves this problem by treating $\Delta N_*$, which relates inflationary scales to observable scales, as a variable parameter. Hence, we could obtain constraints on parameter space by marginalising $\Delta N_*$ over its allowed range, although we do not consider the concrete mechanism to end DBI inflation. In future analysis, more reliable constraints on the parameter space should be obtained with the same methodology but by scanning over a reasonably broad parameter space and then marginalising parameters relevant to the details of the reheating process. 

Furthermore, we make a power spectrum analysis to reduce the number of scenarios before bispectrum analysis, which Clarke does not include. In his $\beta$-only parameter scan, scenarios with $\beta < 0.22$ and $\beta > 0.39$ are disfavoured by the amplitude of scalar power spectrum (model predictions are outside of the 1$\sigma$ region of Planck values), which are ruled out before being forwarded to bispectrum analysis in our work. This discovery further illustrates the $\beta$ dependence of the scalar power spectrum, though less obvious than the $\lambda$ and $V_0$ dependence. Therefore, we conclude that observational results of the scalar power spectrum can impose constraints on $\beta$ as well, but the constraints are loose and sensitively depend on other parameters. We leave this as future work and mainly focus on the marginalised constraint from the bispectrum analysis in this paper.

To confront theoretical predictions with observation, Clarke connects \texttt{Primodal} to the \texttt{CMB-BEST} estimator \cite{Sohn:2023fte} and use Planck 2018 temperature data only, while we connect \texttt{Primodal} to the \texttt{Modal} estimator and use Planck 2018 T+E data. Unlike the \texttt{Modal} estimator, which has a different working basis for the primordial bispectra and CMB bispectra, \texttt{CMB-BEST} only makes a mode decomposition for the primordial bispectrum. Afterwards, a \texttt{KSW} approach is applied to each eigenmode to evaluate $f_{\mathrm{NL}}$ or $c_{\mathrm{NL}}$. As a result, \texttt{CMB-BEST} can avoid the discrepancy from basis transformations to evaluate non-Gaussianity more accurately. Although the \texttt{KSW} approach increases the complexity of the calculation by about 300 times, depending on the choice of basis \cite{Sohn:2023fte}, such calculation still only needs to be paid once per basis. Once the \texttt{KSW} evaluation for a specific basis is done, \texttt{CMB-BEST} can work as efficiently as \texttt{Modal} for bispectrum decomposed w.r.t. that specific basis. The agreement between \texttt{Modal} and \texttt{CMB-BEST} while constraining standard templates is examined and discussed in \cite{Sohn:2023fte}. In the future, we would like to implement the \texttt{Primodal} scaling basis to \texttt{CMB-BEST} and use it as the late-time estimator for $c_{\mathrm{NL}}$ in our pipeline. With extra computational costs, this may effectively avoid the inaccuracy introduced by basis transformations and potentially works better for some targetted subspaces.

The inclusion of power spectrum analysis and E-mode polarisations is the main reason we can get a tighter constraint on $c_s$ even if we scan over more scenarios by allowing all parameters to vary.  

Equation (56) in Planck 2018: Primordial Non-Gaussianity \cite{Planck:2019kim} presents the constraint on the sound speed of DBI inflation from CMB T+E data,
\begin{equation}
    c_s \geq 0.086 \qquad (95\%,\quad T+E,\quad Planck),
\label{7E5}
\end{equation}
using the relationship (\ref{4E5}) and the scale-independent template (\ref{4E4}) (And the constraint on $\beta$ from the Planck 2013 T data is $\beta < 0.7$). It is not unexpected that our pipeline doesn't exactly reproduce the Planck constraint. As discussed in \cite{Clarke:2022kvv}, one possible reason is the different scale used in \texttt{Primodal} -- we are limited to $k_{max}/k_{min} = 1000$ to make \texttt{Primodal} calculations converge, so Planck data with scales $k < k_{min} \approx 1.4\times 10^{-4}$ are not included in our work. This effect is almost negligible for the present shape but can be more important when investigating local-type non-Gaussianities.

Another possible reason is the inaccuracy introduced by basis transformations: In our pipeline, two basis transformations are needed--- from the \texttt{Primodal} basis to the Planck primordial basis, then from the projected Planck primordial basis to the Planck CMB basis. Although \texttt{Primodal} calculations of DBI bispectra can achieve good convergence with a \texttt{Primodal} basis of $p_{max} = 30$ \cite{Clarke:2020znk}, which has about 4000 modes, the Planck Primordial basis, currently having 2000 modes, is unlikely to preserve such high accuracy. Before and after this basis transformation, the DBI bispectrum is changed by a relative error of 2\% due to the unphysical wiggles introduced by the Fourier mode functions. Fortunately, most of these wiggles are unobservable with the resolution of Planck data, so we do not expect this to contribute to a large discrepancy. However, while working with higher-resolution data from future experiments such as the Simons, this issue needs to be carefully treated(As discussed above, using \texttt{CMB-BEST} can avoid this problem at additional computational cost).

Most importantly, the Planck results come from the theoretical template of the DBI bispectrum, which only captures leading-order terms in slow-roll. In contrast, our constraints come from the full numerical calculation without any assumption. For the scenario defined from equation (\ref{6E1.1}) to (\ref{6E1.5}), the bispectrum predicted by \texttt{Primodal} has a relative error of 4\% and 0.7\% compared to Planck's DBI template without and with scaling, respectively. Although the difference is slight in this case, our fully numerical pipeline is expected to have a significant advantage when applying to models which don't have any analytic template or whose non-Gaussianities include non-negligible contributions from sub-leading order terms in slow-roll.

Finally, we conclude that our constraints on IR DBI sound speed and $\beta$ validate our conceptual pipeline as a whole. This pipeline efficiently confronts the inflationary Lagrangian to observational data, enabling us to utilise the information from the full bispectrum (use the information from both the shape and the amplitude of the bispectrum) to investigate general inflation models, especially those which do not have standard templates. Another key advantage of our pipeline is the accuracy and convenience of constraining any parameters and quantities relevant to inflation models, not relying on any analytic approximation. Moreover, our pipeline provides an efficient method for exploring the likelihood functions and investigating basic statistics of the inflationary parameter space with bispectra, laying a solid foundation to develop advanced statistical tools for precision data analysis.

Our future plan can be roughly divided into two parts. For the short term, we will use the constraining pipeline to investigate other general single-field inflation models, including those with sharp feature and periodic feature \cite{Chen:2006xjb}\cite{Chen:2008wn}\cite{Chen:2010xka}, and more complicated models without standard analytical templates, such as those motivated by open system\cite{Salcedo:2024smn}, endeavouring to look for new shapes of PNG and constrain them with Planck data. We note that our methodology is specifically targeting the latest data from the Simons Observatory \cite{SimonsObservatory:2018koc}, but can be equally well applied to other forthcoming observations, like LiteBIRD \cite{LiteBIRD:2022cnt} and galaxy surveys, enabling us to investigate complicated models (such as oscillatory models with high frequency), which are unconstrained by Planck templates, with improved precision and polarisation sensitivity.

In the longer term, we will generalise the \texttt{Modal} decomposition approach in \texttt{Primodal} to a wider range of generic inflation scenarios. This requires treatments for the extra complexity introduced by multiple fields, such as the conversion between the curvature field and the isocurvature fields \cite{Wang:2022eop}. The extension to multi-field models allows for a much richer set of bispectrum signatures to be investigated, with the potential to identify observables that may be linked to additional scalar fields, extra dimensions, the cosmological collider, or alternative scenarios of inflation.

\acknowledgments

We acknowledge first the debt that the present work owes to Philip Clarke and all his efforts creating the \texttt{Primodal} pipeline for efficiently evaluating primordial bispectra and this paper follows up his previous indicative work. We are also grateful to Wuhyun Sohn, Petar Suman and Steven Gratton for useful conversations and comments.  We thank Juliana Kwan for guidance on efficient HPC implementation. BZ acknowledges PhD funding from UKRI (STFC). EPS and JF acknowledge support from the STFC Astronomy Consolidated Grant ST/P000673/1.  Numerical simulations were undertaken on the Cambridge CSD3 part of the STFC DiRAC HPC Facility (www.dirac.ac.uk) funded by BEIS capital funding via STFC Capital Grants ST/P002307/1 and ST/R002452/1 and STFC Operations Grant ST/R00689X/1.  This work has also benefitted from UKRI ExCALIBUR funding EP/Y028082/1.

\bibliographystyle{unsrt}
\bibliography{ref.bib}

\end{document}